\documentclass[showpacs,amssymb,preprint,preprintnumbers,nofootinbib,superscriptaddress]{revtex4}
\usepackage{epsf,epsfig,graphics,graphicx}
\usepackage{verbatim,color,ulem}
\usepackage{xcolor}
\usepackage{longtable}
\newcommand{\six}[6]{\left(\begin{array}{ccc}
									{#1}& {#2}& {#3}\\
									{#4}& {#5}& {#6} \\
\end{array}\right)}

\begin{document}

\title{Four-point correlation functions in axion inflation}
\author{Hing-Tong Cho}
\email[Email: ]{htcho@mail.tku.edu.tw}
\affiliation{Department of Physics, Tamkang University, Tamsui, New Taipei City 25137, Taiwan}
\author{Kin-Wang Ng}
\email[Email: ]{nkw@phys.sinica.edu.tw}
\affiliation{Institute of Physics, Academia Sinica, Taipei 11529, Taiwan}
\affiliation{Institute of Astronomy and Astrophysics, Academia Sinica, Taipei 11529, Taiwan}

\date{\today}

\begin{abstract}

We study parity violation in the early universe by examining the four-point correlation function within the axion inflation model. Using an open quantum system formalism from our previous work, we calculate the influence functional to fourth order, from which we then derive the inflaton four-point correlation function. When we decompose this function using isotropic basis functions, the expansion coefficients $\zeta_{\ell',\ell'',\ell'''}$ naturally split into parity-even and parity-odd components. In the large $\xi$ approximation, which enhances the production of right-handed photons in the model, the derivation of these coefficients simplifies. We work out the lowest-order nonvanishing parity-odd $\zeta_{234}$ term which clearly indicates the presence of parity violation. Moreover, our derived values of the coefficients are 
consistent with recent observational data from galaxy surveys.
\end{abstract}

\maketitle

\section{Introduction}
Identifying parity violation in cosmic structures could provide crucial insights into physics beyond the standard cosmological model. While the $\Lambda$CDM model \cite{LL20} has been remarkably successful, it presumes that gravitational interactions are inherently parity symmetric. Observing parity violation would point to new phenomena in the early universe and refine our understanding of cosmic evolution. Recent galaxy distribution measurements \cite{PHS21,HSC23,Philcox22} indicate possible evidence of such parity-breaking effects, although it was not found at the largest cosmological scales probed by cosmic microwave background (CMB) measurements \cite{Philcox25}.

A well-studied candidate for introducing parity violation in the early universe is the axion inflation model \cite{FFO90,ABFFO93}. The model employs a Chern-Simons coupling between the axion and photon, which naturally generates a parity-violating component. Initially, this coupling was proposed to create a flat potential to address the fine-tuning problem. Later, it was found to have other interesting implications, such as producing non-Gaussian and non-scale-invariant density power spectra \cite{BP11,MP13,ChoNg20} and generating chiral gravitational waves \cite{AS12}.

Parity-violating signatures in axion inflation appear in four-point or higher-point correlation functions. The four-point correlation function (4PCF), which examines the probability of finding quartets of galaxies in specific configurations, is particularly sensitive to parity-violating features. However, its evaluation involves complicated multiple integrals, especially the angular ones with combinations of polarization vectors. To address these challenges, advanced numerical methods have been developed \cite{NRSX23,FMOS24,RSHG24,BWXZ25}. Yet, these methods have limitations in precision and also flexibility in exploring different parts of the parameter-space. To overcome these issues, Cahn and collaborators introduced isotropic basis functions \cite{CS23,HCPS22}, constructed from Wigner-3j symbols, to analytically handle angular integrations, leaving radial equations to be solved numerically \cite{RSHG24}.

In earlier work \cite{ChoNg20}, we investigated the axion inflation model using the closed-time-path \cite{Jordan86,CH87} or Schwinger-Keldysh formalism \cite{Schwinger61,Keldysh65}, along with the influence functional method \cite{FV63,CH08}, to derive in-in correlation functions of the axion-inflaton field. The axion-photon coupling leads to the generation of photons with a single polarization, causing parity violation. In this open quantum system approach, the electromagnetic field is traced out and the parity-breaking effect will reside in the resulting influence functional, which includes noise and dissipation terms. By associating the noise term with a stochastic field linearly coupled to the inflaton, we systematically derived various in-in correlation functions.

Our previous study \cite{ChoNg20} focused on the two- and three-point correlation functions to explore non-Gaussianity in the power spectrum. Here, we extend our analysis to the 4PCF, which requires evaluating the noise kernel to higher orders. The resulting expression involves angular integrations with integrands of complicated combinations of photon polarization vectors, which can be simplified using isotropic basis functions in momentum space \cite{CS23,HCPS22}. These functions also facilitate the decomposition of the 4PCF again using the isotropic basis functions, but now in configuration space. The coefficients of this decomposition can then be used to extract information on parity-violating effects \cite{Philcox22}.

The next section summarizes the influence functional formalism from our earlier work \cite{ChoNg20}. Higher-order noise kernel evaluations are detailed in Appendix A. Section III discusses the derivation of the 4PCF through functional derivatives of the effective action, with further details provided in Appendix B. In Section IV, we decompose the 4PCF using isotropic basis functions and explicitly compute the corresponding coefficients, clarifying their roles in detecting parity-violating effects. In Section V, we discuss the implications of the 4PCF measurements on the axion inflation model. The paper concludes with discussions in Section VI. Some properties of the Wigner-3j symbols are detailed in Appendix C. The radial and angular integrations needed for the evaluation of the coefficients in the isotropic basis function expansion of the 4PCF in Section IV are explicitly worked out in Appendix D. 
A toy inflation model consistent at all relevant scales from CMB scale down to the galaxy scale is provided in Appendix E.

\section{Influence action of axion inflation}
Here we recapitulate the derivation of the influence action of the axion inflation in our earlier work \cite{ChoNg20} (also see the references therein) and extend to the 4th order perturbation.
 We start with the following action of axion inflation consisting of the inflaton field $\varphi (x)$, the photon field $A_{\mu}(x)$, and their interaction,
\begin{equation}
S_{tot}=\int d^{4} x\sqrt{-g}  \left[\frac{M_{P}^{2} }{{2}} R-\frac{1}{2} \left(\partial \varphi \right)^{2} -V(\varphi )-\frac{1}{4} F^{\mu \nu } F_{\mu \nu } -\frac{\alpha }{4f} \varphi \, \tilde{F}^{\mu \nu } F_{\mu \nu } \right],
\label{Stot}
\end{equation}
where $R$ is the curvature scalar, $M_{P}$ the reduced Planck mass, $V(\varphi)$ the inflaton potential,
$f$ the axion decay constant, and $\alpha $ is a dimensionless parameter. The electromagnetic field is $F_{\mu \nu } =\partial _{\mu } A_{\nu } -\partial _{\nu } A_{\mu } $ and $\tilde{F}^{\mu\nu} = \frac{1}{2}\epsilon^{\mu\nu\alpha\beta} F_{\alpha\beta}/{\sqrt{-g}}$ is its dual, where the potential $A_{\mu }$ satisfies the gauge conditions: $A_{\tau} =0$ and $\nabla ^{\mu } A_{\mu } =0$. Then one can write $A_{\mu } =(0,\vec{A})$ and expand $\vec{A}$ in terms of the two physical degrees of freedom,
\begin{equation}
\vec{A} (\vec{x},\tau )=\sum _{\lambda =R,L}\int \frac{d^{3} k}{(2\pi )^{3/2} }  {\kern 1pt} \, e^{i\vec{k}\cdot \vec{x}} A_{\lambda } (\vec{k},\tau )\hat{\varepsilon }_{\lambda } (\hat{k}),
\end{equation}
where $\hat{\varepsilon }_{\lambda } (\hat{k})$ are the circular polarization vectors.

The inflaton field can be separated into a homogeneous background and its fluctuations, $\varphi (\vec{x},\tau )=\Phi (\tau )+\phi (\vec{x},\tau )$. The homogeneous part induces a period of inflation in which we assume the spacetime to be conformally flat with the metric $ds^{2} =a^{2} (\tau )(-d\tau ^{2} +d\vec{x}\cdot d\vec{x})$. The cosmic scale $a(\tau)=-1/H\tau$, where $\tau$ is the conformal time and $H=(da/d\tau)/a^{2}$ is the Hubble parameter. Due to the coupling with $\Phi(\tau)$, one of the two polarizations of the photon field, the right-handed $A_{R}(\vec{k},\tau)$, will be enhanced while the other polarization $A_{L}(\vec{k},\tau)$ will not. This happens when the parameter $\xi=(\alpha\,d\Phi/dt)/2fH$ is large enough with $t$ being the cosmic time defined by $dt=a(\tau)d\tau$. Therefore, we shall retain only $A_{R}(\vec{k},\tau)$, treating it as a scalar field $A(\vec{x},\tau)$. 

According to Eq.~(\ref{Stot}), the action for this $A(\vec{x},\tau)$ can be expressed as
\begin{eqnarray}
S_{0} [A]+S_{int} [A,\phi ]
&=&-\frac{1}{2} \int d\tau \int d^{3} k{\kern 1pt} {\kern 1pt} {\kern 1pt} A^{*} (\vec{k},\tau )\left[\partial _{\tau}^{2} +\vec{k}^{2} - 2aH\xi |\vec{k}| \right] A(\vec{k},\tau )\nonumber\\
&&-\frac{\alpha }{f} \int d^{4} x{\kern 1pt} {\kern 1pt} d^{4} x'd^{4} x''\; A(x)A(x')\phi (x'')\mathcal{F} (x,x',x''),
\end{eqnarray}
where the Fourier transform of $A(\vec{x},\tau)$ in the large $\xi$ limit,
\begin{eqnarray}\label{Amode}
A(k,\tau) &\sim& \frac{1}{\sqrt{2k}}\left(\frac{k|\tau|}{2\xi}\right)^{1/4} e^{\pi\xi-2\sqrt{2k|\tau|\xi}},
\end{eqnarray}
and
\begin{eqnarray}\label{Fint}
\mathcal{F} (x,x',x'')&=&\delta(\tau-\tau'') \partial_{\tau'}\delta(\tau'-\tau'')\nonumber\\
&&\ \ \int \frac{d^{3}k}{(2\pi )^{3}} \frac{d^{3}k'}{(2\pi )^{3}} {\kern 1pt} \, e^{-i\vec{k}\cdot (\vec{x}-\vec{x}'')} \, e^{-i\vec{k}'\cdot (\vec{x}'-\vec{x}'')}
|\vec{k}|(\hat{\varepsilon }_{R} (\hat{k})\cdot \hat{\varepsilon }_{R} (\hat{k}')) {\rm \; }.
\end{eqnarray}

To implement the in-in or the closed-time-path (CTP) formalism, we use two copies of both the inflaton and the photon fields. The corresponding action can be written as
\begin{eqnarray}
S=S_{0} [A_{+} ]-S_{0} [A_{-} ]+S_{int} [A_{+} ,\phi _{+} ]-S_{int} [A_{-} ,\phi _{-} ].
\end{eqnarray}
The influence of the photon field on the inflaton is summarized by the so-called influence action $S_{\rm IF}$ defined by integrating over the photon degrees of freedom in the path integrals,
\begin{eqnarray}
e^{iS_{IF} } =\int _{CTP}DA_{+} DA_{-}  e^{iS},
\label{SIF}
\end{eqnarray}
with the CTP boundary conditions. A perturbative series for $S_{IF}$ can be developed by adding sources to the photon action,
\begin{eqnarray}\label{SIFj}
e^{iS_{IF} } &=&\left.\int _{CTP}DA_{+} DA_{-}  e^{iS+i\int J_{+} A_{+}  -i\int J_{-} A_{-}  }\right|_{J_{+}=J_{-}=0}\nonumber\\
&=&e^{iS_{int} \left[\frac{1}{i} \frac{\delta }{\delta J_{+} } ,\phi _{+} \right]-iS_{int} \left[\frac{1}{-i} \frac{\delta }{\delta J_{-} } ,\phi _{-} \right]} \left.\int _{CTP}DA_{+} DA_{-}  e^{i\left(S_{0} [A_{+} ]-S_{0} [A_{-} ]+\int J_{+} A_{+} -\int J_{-} A_{-}   \right)}\right|_{J_{+}=J{-}=0}\nonumber\\
&=&\left. e^{iS_{int} \left[\frac{1}{i} \frac{\delta }{\delta J_{+} } ,\phi _{+} \right]-iS_{int} \left[\frac{1}{-i} \frac{\delta }{\delta J_{-} } ,\phi _{-} \right]}e^{-\frac{i}{2} \int \left(J_{+} G_{++} J_{+} -J_{+} G_{+-} J_{-} -J_{-} G_{-+} J_{+} +J_{-} G_{--} J_{-} \right) }\right|_{J_{+}=J{-}=0},\nonumber\\
\end{eqnarray}
where we have the Schwinger-Keldysh Green functions, $G_{++}$, $G_{+-}$, $G_{-+}$, and $G_{--}$, due to the CTP boundary conditions. In the large $\xi$ approximation, the mode function of the photon field is given by Eq.~(\ref{Amode}) which is real and depends only on the magnitude $k$. 
Therefore, all the Schwinger-Keldysh Green functions are equal to
\begin{eqnarray}\label{gfct}
G(x,x')&=&-i\int \frac{d^{3} k}{(2\pi )^{3} }  {\kern 1pt} \, A(k,\tau) A(k,\tau'){\kern 1pt} {\kern 1pt} e^{i\vec{k}\cdot (\vec{x}-\vec{x}')}\nonumber\\
&=&\int\,\frac{d^{3}k}{(2\pi)^{3}}\,e^{i\vec{k}\cdot(\vec{x}-\vec{x}')}g_{k}(\tau,\tau'),
\end{eqnarray}
where
\begin{eqnarray}\label{gk}
g_{k}(\tau,\tau')=-ie^{2\pi\xi}\frac{(\tau\tau')^{1/4}}{2\sqrt{2\xi k}}e^{-2\sqrt{2\xi k}(\sqrt{-\tau}+\sqrt{-\tau'})}\,.
\end{eqnarray}
This would happen if the photon field becomes tachyonic and the inflaton decays into photons dissipationlessly.
Thus, the absence of the dissipation kernel will simplify significantly
the computation of the correlation functions \cite{ChoNg20}.

Next, we shall expand the influence action $S_{IF}$ as a power series in $\alpha/f$,
\begin{eqnarray}
S_{IF}=\sum_{i=1}^{\infty}\delta S^{(i)}_{IF}.
\end{eqnarray}
This is done by taking functional derivatives of $J_{+}$ and $J_{-}$ in Eq.~(\ref{SIFj}). The terms $\delta S_{IF}^{(i)}$, for $i=1,2,3$, have been worked out in Ref.~\cite{ChoNg20}. Let us summarize them here.
\begin{eqnarray}
\delta S_{IF}^{(1)}=-\int d^4x \sqrt{-g(x)}\, \delta V(x) \Delta\phi(x), \nonumber\\
\end{eqnarray}
where
\begin{eqnarray}
\delta V(x)= \frac{1}{a(\tau)} \left(\frac{\alpha}{f}e^{2\pi\xi}\right) \left(\frac{135}{65536\pi^2}\right) \frac{1}{\xi^4\tau^4}
\end{eqnarray}
and $\Delta\phi(x)\equiv\phi_{+}(x)-\phi_{-}(x)$.
\begin{eqnarray}
\delta S_{IF}^{(2)}={i\over 2} \int d^4x_1 d^4x_2 \sqrt{-g(x_1)} \sqrt{-g(x_2)} \Delta\phi(x_1) N_2(x_1,x_2) \Delta\phi(x_2), \nonumber
\end{eqnarray}
where
\begin{eqnarray}
&&N_2(x_1,x_2) \nonumber\\
&=&  \frac{1}{a^4(\tau_1)a^4(\tau_2)} \left(\frac{\alpha^2}{f^2}e^{4\pi\xi}\right)
\int\frac{d^{3}k}{(2\pi)^{3}}e^{i\vec{k}\cdot(\vec{x}_{1}-\vec{x}_{2})}e^{-2\sqrt{2\xi k}(\sqrt{-\tau_{1}}+\sqrt{-\tau_{2}})}\nonumber\\
&&\Bigg\{\left[\left(\frac{15\,k^{3/2}}{256\sqrt{2}\,\pi^{2}\,\xi^{7/2}}\right)\frac{1}{(\sqrt{-\tau_{1}}+\sqrt{-\tau_{2}})^{7}}+\left(\frac{105\,k}{1024\,\pi^{2}\,\xi^{4}}\right) \frac{1}{(\sqrt{-\tau_{1}}+\sqrt{-\tau_{2}})^{8}}\right]+\dots\Bigg\}\,.\nonumber\\
\end{eqnarray}
\begin{eqnarray}
\delta S_{IF}^{(3)}={1\over 6} \int d^4x_1 d^4x_2 d^4x_3 \sqrt{-g(x_1)} \sqrt{-g(x_2)} \sqrt{-g(x_3)}\Delta\phi(x_1)\Delta\phi(x_2)\Delta\phi(x_3) N_3(x_1,x_2,x_3),\nonumber\\
\end{eqnarray}
where
\begin{eqnarray}
&&N_3(x_1,x_2,x_3)\nonumber\\
&=& \frac{1}{a^4(\tau_1)a^4(\tau_2)a^4(\tau_3)}
\left(\frac{\alpha^{3}}{f^{3}}e^{6\pi\xi}\right) \times\nonumber\\
&&\int\frac{d^{3}k}{(2\pi)^{3}}\int\frac{d^{3}k'}{(2\pi)^{3}}\,e^{i\vec{k}\cdot(\vec{x}_{1}-\vec{x}_{2})}\,e^{i\vec{k}'\cdot(\vec{x}_{2}-\vec{x}_{3})}
\times\nonumber\\
&&\Bigg\{\left[ \frac{105\,(k^{3/2}k'^{1/2}+kk'/3)}{8192\pi^2\xi^{4}}\right]\left(1+\frac{\vec{k}\cdot\vec{k}'}{kk'}\right)^{2}\
\frac{e^{-2\sqrt{2\xi}[(\sqrt{-\tau_{1}}+\sqrt{-\tau_{2}})\sqrt{k}+(\sqrt{-\tau_{2}}+\sqrt{-\tau_{3}})\sqrt{k'}]}}{(\sqrt{-\tau_1}+\sqrt{-\tau_3})^{8}}\nonumber\\
&&\ \ +\,5\ {\rm permutations\ of\ }(x_{1},x_{2},x_{3})+ \dots\Bigg\}\,.
\end{eqnarray}

In order to consider the 4PCF of the inflaton field in the following section, we shall see that we need to evaluate $\delta S_{IF}^{(4)}$. From Eq.~(\ref{SIFj}), this term involves the following functional derivatives,
\begin{eqnarray}
&&\frac{\alpha^4}{24f^4}\Bigg\{i\int d^{4} x{\kern 1pt} {\kern 1pt} d^{4} x'd^{4} x''\; \left[\frac{1}{i} \frac{\delta }{\delta J_{+}(x)} \frac{1}{i} \frac{\delta }{\delta J_{+} (x')} \phi_{+} (x'')- \frac{1}{-i} \frac{\delta }{\delta J_{-}(x)} \frac{1}{-i} \frac{\delta }{\delta J_{-} (x')} \phi_{-} (x'')\right] \nonumber \\
&&\left.  \mathcal{F} (x,x',x') \
\Bigg\}^4\
e^{-\frac{i}{2} \int \left(J_{+} G_{++} J_{+} -J_{+} G_{+-} J_{-} -J_{-} G_{-+} J_{+} +J_{-} G_{--} J_{-} \right) } \right|_{J_{+} =J_{-} =0}.
\end{eqnarray}
Since in the large $\xi$ approximation all the Schwinger-Keldsyh Green's functions equal to $G(x,x')$ given in Eq.~(\ref{gfct}), the resulting expression for the functional derivatives above contains only $\Delta\phi(x)$:
\begin{eqnarray}\label{allterms}
&&\frac{\alpha^{4}}{24f^{4}} \int\,d^{4}x_{1}\cdots d^{4}x_{12}\ \Delta\phi(x_{9})\Delta\phi(x_{10})\Delta\phi(x_{11})\Delta\phi(x_{12})\nonumber\\
&&\hskip 40pt \mathcal{F}(x_{1},x_{2},x_{9})\mathcal{F}(x_{3},x_{4},x_{10})\mathcal{F}(x_{5},x_{6},x_{11})\mathcal{F}(x_{7},x_{8},x_{12}) \nonumber \\
&&\hskip 38pt \bigg\{\big[G(x_{1},x_{3})G(x_{2},x_{5})G(x_{4},x_{7})G(x_{6},x_{8})+G(x_{1},x_{3})G(x_{2},x_{5})G(x_{4},x_{8})G(x_{6},x_{7}) \nonumber \\
&&\hskip 50pt G(x_{1},x_{3})G(x_{2},x_{6})G(x_{4},x_{7})G(x_{5},x_{8})+G(x_{1},x_{3})G(x_{2},x_{6})G(x_{4},x_{8})G(x_{5},x_{7}) \nonumber \\
&&\hskip 50pt G(x_{1},x_{3})G(x_{2},x_{7})G(x_{4},x_{5})G(x_{6},x_{8})+G(x_{1},x_{3})G(x_{2},x_{7})G(x_{4},x_{6})G(x_{5},x_{8}) \nonumber \\
&&\hskip 50pt G(x_{1},x_{3})G(x_{2},x_{8})G(x_{4},x_{5})G(x_{6},x_{7})+G(x_{1},x_{3})G(x_{2},x_{8})G(x_{4},x_{6})G(x_{5},x_{7})\big] \nonumber \\ 
&&\hskip 55pt +(3\leftrightarrow 4)+(3\leftrightarrow 5)+(3\leftrightarrow 6)+(3\leftrightarrow 7)+(3\leftrightarrow 8) \bigg\}\nonumber\\
&&\hskip 100pt +\;{\rm disconnected\ terms}\,.
\end{eqnarray}
Ignoring disconnected terms, there are 48 connected ones to be evaluated in total. From their symmetries, these 48 terms can be grouped into 3 types. In the first type, the coordinates $x_{1}$, $x_{3}$, $x_{5}$, and $x_{7}$ are distributed in only two Green's functions. There are 6 terms of this type. In the second type, the four coordinates are distributed in three Green's functions, and there are 36 of them. Finally, in the third type, the four coordinates are distributed in four different Green's functions, and there are 6 of them. 

To consider the terms of the first type, we look at a typical one with integrations,
\begin{eqnarray}\label{type1}
&& \int\,d^{4}x_{1}\cdots d^{4}x_{8}\ \mathcal{F}(x_{1},x_{2},x_{9})\mathcal{F}(x_{3},x_{4},x_{10})\mathcal{F}(x_{5},x_{6},x_{11})\mathcal{F}(x_{7},x_{8},x_{12}) \nonumber \\
&&\hskip 38pt G(x_{1},x_{3})G(x_{2},x_{8})G(x_{4},x_{6})G(x_{5},x_{7}) \nonumber\\
&=&\int\frac{d^{3}K}{(2\pi)^{3}}\frac{d^{3}K'}{(2\pi)^{3}}\frac{d^{3}K''}{(2\pi)^{3}}\,e^{i\vec{K}\cdot(\vec{x}_{9}-\vec{x}_{10})}e^{i\vec{K}'\cdot(\vec{x}_{10}-\vec{x}_{11})}e^{i\vec{K}''\cdot(\vec{x}_{11}-\vec{x}_{12})}\nonumber\\
&&\ \ K^{3/2}K'^{1/2}K''^{3/2}\,(\hat{\varepsilon}_{R}(\hat{K})\cdot\hat{\varepsilon}_{R}(-\hat{K}''))(\hat{\varepsilon}_{R}(\hat{K}'')\cdot\hat{\varepsilon}_{R}(-\hat{K}'))(\hat{\varepsilon}_{R}(\hat{K}')\cdot\hat{\varepsilon}_{R}(-\hat{K}))\nonumber\\
&&\ \ \left(\frac{e^{8\pi\xi}}{\xi^{7/2}}\right)\left(\frac{15\sqrt{2}}{2048\pi^{2}}\right)\frac{1}{(\sqrt{-\tau_{9}}+\sqrt{-\tau_{12}})^{7}}\nonumber\\
&&\ \ e^{-2\sqrt{2\xi K}(\sqrt{-\tau_{9}}+\sqrt{-\tau_{10}})}e^{-2\sqrt{2\xi K'}(\sqrt{-\tau_{10}}+\sqrt{-\tau_{11}})}e^{-2\sqrt{2\xi K''}(\sqrt{-\tau_{11}}+\sqrt{-\tau_{12}})}\,.
\end{eqnarray}
The details of the calculation have been laid out in Appendix \ref{appsec:noise}. The other five terms in this type can be obtained by permuting the coordinates $x_{9}$, $x_{10}$, $x_{11}$, and $x_{12}$ in Eq.~(\ref{type1}).

For the second type of terms, with the coordinates $x_{1}$, $x_{3}$, $x_{5}$, and $x_{7}$ in three different Green's functions, the result of the evaluation of a typical one would be
\begin{eqnarray}\label{type2}
&& \int\,d^{4}x_{1}\cdots d^{4}x_{8}\ \mathcal{F}(x_{1},x_{2},x_{9})\mathcal{F}(x_{3},x_{4},x_{10})\mathcal{F}(x_{5},x_{6},x_{11})\mathcal{F}(x_{7},x_{8},x_{12}) \nonumber \\
&&\hskip 38pt G(x_{1},x_{3})G(x_{2},x_{8})G(x_{4},x_{5})G(x_{6},x_{7}) \nonumber\\
&=&\int\frac{d^{3}K}{(2\pi)^{3}}\frac{d^{3}K'}{(2\pi)^{3}}\frac{d^{3}K''}{(2\pi)^{3}}\,e^{i\vec{K}\cdot(\vec{x}_{9}-\vec{x}_{10})}e^{i\vec{K}'\cdot(\vec{x}_{10}-\vec{x}_{11})}e^{i\vec{K}''\cdot(\vec{x}_{11}-\vec{x}_{12})}\nonumber\\
&&\ \ K^{3/2}K'K''\,(\hat{\varepsilon}_{R}(\hat{K})\cdot\hat{\varepsilon}_{R}(-\hat{K}''))(\hat{\varepsilon}_{R}(\hat{K}'')\cdot\hat{\varepsilon}_{R}(-\hat{K}'))(\hat{\varepsilon}_{R}(\hat{K}')\cdot\hat{\varepsilon}_{R}(-\hat{K}))\nonumber\\
&&\ \ \left(\frac{e^{8\pi\xi}}{\xi^{7/2}}\right)\left(\frac{15\sqrt{2}}{2048\pi^{2}}\right)\frac{1}{(\sqrt{-\tau_{9}}+\sqrt{-\tau_{12}})^{7}}\nonumber\\
&&\ \ e^{-2\sqrt{2\xi K}(\sqrt{-\tau_{9}}+\sqrt{-\tau_{10}})}e^{-2\sqrt{2\xi K'}(\sqrt{-\tau_{10}}+\sqrt{-\tau_{11}})}e^{-2\sqrt{2\xi K''}(\sqrt{-\tau_{11}}+\sqrt{-\tau_{12}})}\,.
\end{eqnarray}
The details of the evaluation of this term are also given in Appendix \ref{appsec:noise}. Again, the other 35 terms in this type can be obtained by permuting the coordinates $x_{9}$, $x_{10}$, $x_{11}$, and $x_{12}$.

Finally, we come to the last type of terms with the coordinates $x_{1}$, $x_{3}$, $x_{5}$, and $x_{7}$ distributed in four different Green's functions. The resulting expression for the typical term in this type is 
\begin{eqnarray}\label{type3}
&& \int\,d^{4}x_{1}\cdots d^{4}x_{8}\ \mathcal{F}(x_{1},x_{2},x_{9})\mathcal{F}(x_{3},x_{4},x_{10})\mathcal{F}(x_{5},x_{6},x_{11})\mathcal{F}(x_{7},x_{8},x_{12}) \nonumber \\
&&\hskip 38pt G(x_{1},x_{4})G(x_{2},x_{7})G(x_{3},x_{6})G(x_{5},x_{8}) \nonumber\\
&=&\int\frac{d^{3}K}{(2\pi)^{3}}\frac{d^{3}K'}{(2\pi)^{3}}\frac{d^{3}K''}{(2\pi)^{3}}\,e^{i\vec{K}\cdot(\vec{x}_{9}-\vec{x}_{10})}e^{i\vec{K}'\cdot(\vec{x}_{10}-\vec{x}_{11})}e^{i\vec{K}''\cdot(\vec{x}_{11}-\vec{x}_{12})}\nonumber\\
&&\ \ KK'K''\,(\hat{\varepsilon}_{R}(\hat{K})\cdot\hat{\varepsilon}_{R}(-\hat{K}''))(\hat{\varepsilon}_{R}(\hat{K}'')\cdot\hat{\varepsilon}_{R}(-\hat{K}'))(\hat{\varepsilon}_{R}(\hat{K}')\cdot\hat{\varepsilon}_{R}(-\hat{K}))\nonumber\\
&&\ \ \left(\frac{e^{8\pi\xi}}{\xi^{4}}\right)\left(\frac{105}{4096\pi^{2}}\right)\frac{1}{(\sqrt{-\tau_{9}}+\sqrt{-\tau_{12}})^{8}}\nonumber\\
&&\ \ e^{-2\sqrt{2\xi K}(\sqrt{-\tau_{9}}+\sqrt{-\tau_{10}})}e^{-2\sqrt{2\xi K'}(\sqrt{-\tau_{10}}+\sqrt{-\tau_{11}})}e^{-2\sqrt{2\xi K''}(\sqrt{-\tau_{11}}+\sqrt{-\tau_{12}})}\,.
\end{eqnarray}
We refer to Appendix \ref{appsec:noise} again for details. By permuting the coordinates $x_{9}$, $x_{10}$, $x_{11}$, and $x_{12}$, one can obtain the other 5 terms in this type.

Now we have basically evaluated all the connected terms in Eq.~(\ref{allterms}) which will contribute to the 4th order influence action $\delta S_{IF}^{(4)}$. Eq.~(\ref{type1}) gives the result of the typical one in the first type of terms. To obtain the total contribution from all the 6 terms in this type, we just need to permute $x_{9}$ to $x_{12}$ and to divide the resulting expression by 4. Similar procedure can be implemented to the other two types. For the second type of terms, we would again permute $x_{9}$ to $x_{12}$ in the expression in Eq.~(\ref{type2}) and then multiply the result by $3/2$ to get the contribution of all the 24 terms of this type. Finally, for the third type of terms, we permute $x_{9}$ to $x_{12}$ in Eq.~(\ref{type3}), and divide the result by 4 to get the total contribution of all the 6 terms of this type. Collecting all these contributions, one can express 
\begin{eqnarray}
    \delta S_{IF}^{(4)}&=&\frac{1}{24}\int d^4x_1 d^4x_2 d^4x_3 d^4x_4\sqrt{-g(x_1)} \sqrt{-g(x_2)} \sqrt{-g(x_3)} \sqrt{-g(x_4)} \nonumber\\
&&\hskip 50pt \Delta\phi(x_1)\Delta\phi(x_2)\Delta\phi(x_3) \Delta\phi(x_4) N_4(x_1,x_2,x_3,x_4),
\end{eqnarray}
where 
\begin{eqnarray}\label{Noise4}
    &&N_4(x_1,x_2,x_3,x_4)\nonumber\\
    &=&\frac{1}{a^4(\tau_1)a^4(\tau_2)a^4(\tau_3)a^4(\tau_4)}
\left(\frac{\alpha^{4}}{f^{4}}e^{8\pi\xi}\right)\nonumber\\
&&\Bigg\{\int\frac{d^{3}k}{(2\pi)^{3}}\int\frac{d^{3}k'}{(2\pi)^{3}}\int\frac{d^{3}k''}{(2\pi)^{3}}\,e^{i\vec{k}\cdot(\vec{x}_{1}-\vec{x}_{2})}\,e^{i\vec{k}'\cdot(\vec{x}_{2}-\vec{x}_{3})}e^{i\vec{k}''\cdot(\vec{x}_{3}-\vec{x}_{4})}\nonumber\\
&&\ \ \ \ k k' k''(\hat{\varepsilon}_{R}(\hat{k})\cdot\hat{\varepsilon}_{R}(-\hat{k}''))(\hat{\varepsilon}_{R}(\hat{k}'')\cdot\hat{\varepsilon}_{R}(-\hat{k}'))(\hat{\varepsilon}_{R}(\hat{k}')\cdot\hat{\varepsilon}_{R}(-\hat{k}))\nonumber\\
&&\ \ \ \ e^{-2\sqrt{2\xi k}(\sqrt{-\tau_{1}}+\sqrt{-\tau_{2}})}e^{-2\sqrt{2\xi k'}(\sqrt{-\tau_{2}}+\sqrt{-\tau_{3}})}e^{-2\sqrt{2\xi k''}(\sqrt{-\tau_{3}}+\sqrt{-\tau_{4}})}\nonumber\\
&&\ \ \ \ \left[\left( \frac{15\sqrt{2}}{8192\pi^2\xi^{7/2}}\right)
\frac{(6k^{1/2}+k^{1/2}k'^{-1/2}k''^{1/2})}{(\sqrt{-\tau_1}+\sqrt{-\tau_3})^{7}}+\left(\frac{105}{16384\pi^{2}\xi^{4}}\right)\frac{1}{(\sqrt{-\tau_1}+\sqrt{-\tau_3})^{8}}\right]\nonumber\\
&&\hskip 100pt \ +\;23\ {\rm permutations\ of\ }(x_{1},x_{2},x_{3},x_{4})\Bigg\}\,.
\end{eqnarray}

The influence action $S_{IF}$ has a stochastic interpretation analogous to the quantum Brownian motion model in which the terms $\delta S_{IF}^{(i)}$ can be represented as a stochastic force $\zeta$ by
\begin{eqnarray}\label{DefP}
    e^{-\frac{1}{2}\int\Delta\phi \,N_{2}\,\Delta\phi+\frac{i}{6}\int\Delta\phi\,\Delta\phi\,\Delta\phi\, N_{3}+\frac{1}{24}\int\Delta\phi\,\Delta\phi\,\Delta\phi\, \Delta\phi\,N_{4}}\equiv\int\,D\zeta\,P[\zeta]\,e^{-i\int\zeta\Delta \phi}\,,
\end{eqnarray}
where $P[\zeta]$ is a probability distribution functional normalized to $\int D\zeta\,P[\zeta]=1$. Taking functional derivatives with respect to $\Delta\phi$ on both sides of Eq.~(\ref{DefP}) and putting $\Delta\phi$ to zero, we have
\begin{eqnarray}
\langle \zeta(x) \rangle 
&=&\int D\zeta P[\zeta] \zeta(x) =0,  \\
\langle \zeta(x_1)\zeta(x_2) \rangle
&=& \int D\zeta P[\zeta] \zeta(x_1)\zeta(x_2) =N_2(x_1,x_2), \label{N2}\\
\langle \zeta(x_1)\zeta(x_2) \zeta(x_3) \rangle
&=& \int D\zeta P[\zeta] \zeta(x_1)\zeta(x_2)\zeta(x_3)=N_3(x_1,x_2,x_3).\label{N3}\\
\langle \zeta(x_1)\zeta(x_2) \zeta(x_3) \zeta(x_4)\rangle 
&=& \int D\zeta P[\zeta] \zeta(x_1)\zeta(x_2)\zeta(x_3)\zeta(x_4)\nonumber\\
&=&N_4(x_1,x_2,x_3,x_4)+N_2(x_1,x_2)N_2(x_3,x_4)\nonumber\\
&&\ \ +N_2(x_1,x_3)N_2(x_2,x_4)+N_2(x_1,x_4)N_2(x_2,x_3).\label{N4}
\end{eqnarray}
Due to the two-point correlator, $N_{2}(x_1,x_2)$ is usually referred to as the noise kernel. Here, it is important to note that the probability functional $P[\zeta]$ is non-Gaussian. That is why the correlators or the stochastic averages cannot be expressed only in terms of $N_{2}(x_1,x_2)$. These correlators are necessary in the evaluation of the n-point correlation functions of the inflaton field $\phi(x)$. In particular, in evaluating the connected 4PCF of the $\phi(x)$ field in the next section, we need the function $N_{4}(x_1,x_2,x_3,x_4)$ as laid out in Eq.~(\ref{Noise4}).

As mentioned above, we have neglected the subdominant contribution of the dissipation kernel.
The magnitudes of the right-hand photon Fourier mode and the left-hand photon mode have a ratio of the order of $e^{\pi\xi}$. Hence, the 4PCF has a factor of $e^{8\pi\xi}$ related to the noise kernel $N_{4}$ (see next section). Compared to a lower-order contribution of the noise kernel, a factor of $e^{2\pi\xi}$ would be a rough estimate for the neglected part of the correlation function. Moreover, the lowest order of the noise kernel $N_{2}$ has a factor of $e^{2\pi\xi}$, and therefore the dissipation kernel should be without this factor and is highly subdominant.


\section{Four-point correlation functions}
In this section we work on the 4PCF of the inflaton field $\phi(x)$. We go back to the CTP path integral over $\phi_{+}(x)$ and $\phi_{-}(x)$, and we add sources so that the correlation functions can be obtained by taking functional derivatives on these terms. The two-point and three-point correlation functions have been analyzed in Ref.~\cite{ChoNg20}. The two-point correlation function 
gives a correction to the power spectrum of the inflaton de Sitter vacuum fluctuations, while the non-Gaussian three-point one gives rise to the power bispectrum. 

We start with the CTP path integral of the effective action $\Gamma[\phi_{+},\phi_{-}]$, including the influence action in Eq.~(\ref{SIF}) and source terms for $\phi_{+}$ and $\phi_{-}$:
\begin{eqnarray}\label{effaction}
&&e^{i\Gamma[J_+, J_-]} \nonumber \\
&=& \int_{CTP} D\phi_+ D\phi_- \int D\zeta P[\zeta] e^{i\int\left[-{1\over2}(\partial\phi_+)^2-{1\over2}V''(\Phi)\phi_+^2\right] +i\int J_+\phi_+}\nonumber\\
&&\hspace{5cm}e^{
-i\int\left[-{1\over2}(\partial\phi_-)^2-{1\over2}V''(\Phi)\phi_-^2\right] -i\int J_-\phi_-}  e^{-i\int \zeta\Delta\phi}  \nonumber \\
&=& \int D\zeta P[\zeta] e^{-{i\over2}\int\left[(J_+-\zeta)G_{++}^\phi(J_+-\zeta) - (J_+-\zeta)G_{+-}^\phi(J_--\zeta)
- (J_--\zeta)G_{-+}^\phi(J_+-\zeta) + (J_--\zeta)G_{--}^\phi(J_--\zeta) \right]},\nonumber\\
\end{eqnarray}
where $G_{++}^{\phi}(x_1,x_2)$, $G_{+-}^{\phi}(x_1,x_2)$, $G_{-+}^{\phi}(x_1,x_2)$, and $G_{--}^{\phi}(x_1,x_2)$ are the Schwinger-Keldysh Green's functions of the $\phi$ field. 

To investigate the parity effects in this axion inflation model, we need to analyze the 4PCF. As we have done in Ref.~\cite{ChoNg20}, the equal-time four-point function can be obtained by taking functional derivatives on the effective action in Eq.~(\ref{effaction}). The resulting quantity is the time-ordered four-point function expressed in terms of the retarded Green's functions. After taking the appropriate equal-time limits, we can finally arrive at the required four-point function,
\begin{eqnarray}\label{four point}
&&\langle \phi(\tau,\vec{x})\phi(\tau,\vec{x}')\phi(\tau,\vec{x}'')\phi(\tau,\vec{x}''')\rangle \nonumber \\
&=&\frac{1}{\sqrt{-g(x)}} \frac{\delta}{\delta J_+(x)}  \frac{1}{\sqrt{-g(x')}} \frac{\delta}{\delta J_+(x')} \times\nonumber\\
&&\ \ \ \ \ \frac{1}{\sqrt{-g(x'')}} \frac{\delta}{\delta J_+(x'')} \frac{1}{\sqrt{-g(x''')}}\frac{\delta}{\delta J_+(x''')}
e^{i\Gamma[J_+, J_-]} {\Big |}_{J_+=J_-=0,\,\tau>\tau'\rightarrow\tau>\tau''\rightarrow\tau>\tau'''\rightarrow\tau} \nonumber \\
&=& \int d^4x_1 \sqrt{-g(x_1)} \int d^4x_2 \sqrt{-g(x_2)} \int d^4x_3 \sqrt{-g(x_3)}  \int d^4x_4 \sqrt{-g(x_4)} \times \nonumber \\
&&\ \ \ \ \  G_{\rm ret}^\phi(x,x_1) G_{\rm ret}^\phi(x',x_2) G_{\rm ret}^\phi(x'',x_3)G_{\rm ret}^\phi(x''',x_4) N_4(x_1,x_2,x_3,x_4) {\Big |}_{\tau>\tau'\rightarrow\tau>\tau''\rightarrow\tau>\tau'''\rightarrow\tau}\nonumber\\
&&\hskip 100pt +{\rm \ disconnected\  terms}\,,
\end{eqnarray}
where the retarded Green's function $G_{ret}^{\phi}(x,x')=-G_{++}^{\phi}(x,x')+G_{-+}^{\phi}(x,x')=-G_{-+}^{\phi}(x,x')+G_{--}^{\phi}(x,x')$.  

Under quantization, the inflaton field becomes
\begin{eqnarray}
    \phi(x)=\int\,d^{3}k\,[a_{\vec{k}}\varphi_{\vec{k}}(\tau,\vec{x})+a_{\vec{k}}^{\dagger}\varphi_{\vec{k}}^{*}(\tau,\vec{x})],
\end{eqnarray}
where $\varphi_{\vec{k}}(\tau,\vec{x})$ is the mode function. If one chooses the Bunch-Davies vacuum during the inflationary era, the mode function can be represented by
\begin{eqnarray}
\varphi_{\vec{k}}(\tau,\vec{x})=\left(-\frac{H}{2^{5/2}\pi\,k^{3/2}}\right)(-k\tau)^{3/2}H_{3/2}^{(1)}(-k\tau)\,e^{i\vec{k}\cdot\vec{x}}.
\label{modefunction}
\end{eqnarray}
Moreover, the retarded Green's function of the inflaton field can be expressed in terms of this mode function as
\begin{eqnarray}\label{retG}
   G_{\rm ret}^{\phi}(x,x')&=&i\theta(\tau-\tau')\int d^{3}k\ \left[\varphi_{\vec{k}}(\tau,\vec{x})\varphi_{\vec{k}}^{*}(\tau',\vec{x}')-\varphi_{\vec{k}}^{*}(\tau,\vec{x})\varphi_{\vec{k}}(\tau',\vec{x}')\right].
\end{eqnarray}
With this retarded Green's function and the expression for $N_{4}(x,x',x'',x''')$ in Eq.~(\ref{Noise4}), the connected part of the 4PCF can be simplified to
\begin{eqnarray}\label{4ptcf1}
&&\left\langle\phi(\tau,\vec{x})\phi(\tau,\vec{x}')\phi(\tau,\vec{x}'')\phi(\tau,\vec{x}''')\right\rangle_{conn}\nonumber\\
    &=&\left(\frac{15}{2^{26}\pi^{2}}\right)\left(\frac{\alpha^{4}}{f^{4}}e^{8\pi\xi}\right)\left(\frac{H(-\tau)}{\xi}\right)^{8}\nonumber\\
    &&\int\frac{d^{3}k}{(2\pi)^{3}}\int\frac{d^{3}k'}{(2\pi)^{3}}\int\frac{d^{3}k''}{(2\pi)^{3}}
    (\hat{\epsilon}_{R}(\hat{k})\cdot\hat{\epsilon}_{R}(-\hat{k}''))
    (\hat{\epsilon}_{R}(\hat{k}'')\cdot\hat{\epsilon}_{R}(-\hat{k}'))(\hat{\epsilon}_{R}(\hat{k}')\cdot\hat{\epsilon}_{R}(-\hat{k}))\nonumber\\
    &&\ \ \left[\frac{e^{-4\sqrt{2\xi(-\tau)}(\sqrt{k}+\sqrt{k'}+\sqrt{k''})}}{(\sqrt{k}+\sqrt{k'})^{2}(\sqrt{k'}+\sqrt{k''})^{2}}\right]
    \Big[7k'+4\sqrt{2\xi(-\tau)}(k^{1/2}k'^{1/2})(6k'^{1/2}+k''^{1/2})\Big]\nonumber\\
    &&\ \ \left\{e^{i\vec{k}\cdot\vec{x}}e^{-i(\vec{k}-\vec{k}')\cdot\vec{x}'}e^{-i(\vec{k}'-\vec{k}'')\cdot\vec{x}''}e^{-i\vec{k}''\cdot\vec{x}''}+23\ {\rm permutations\ of }\ \vec{x},\vec{x}',\vec{x}'',\vec{x}'''\right\}\,.
\end{eqnarray}
The derivation of this result is detailed in Appendix~\ref{appsec:der4pt}.

The $\vec{k}$ integrals in this connected correlation function can be done with the help of the spherical wave expansion formula,
\begin{eqnarray}\label{eexp}
    e^{i\vec{k}\cdot\vec{x}}=\sum_{\ell m}4\pi i^{\ell }j_{\ell }(kx)Y_{\ell m}^{*}(\hat{k})Y_{\ell m}(\hat{x}).
\end{eqnarray}
Hence, the integrations over $k$'s will involve $j_{\ell }(kx)$. Due to the $e^{-4\sqrt{2\xi(-\tau)k}}$ factor, the dominant part of the $k$ integration comes from small values of $k$ and one can approximate
\begin{eqnarray}
    j_{\ell }(kx)=\frac{(kx)^{\ell }}{(2\ell +1)!!}+\cdots
\end{eqnarray}
Then, in the large $\xi$ approximation, the integrations over the magnitudes $k$, $k'$, and $k''$ can be represented by the function,
\begin{eqnarray}\label{calR}
   && {\cal R}_{\ell _{1}\cdots \ell _{6}}\nonumber\\
   &=&\prod_{i=1}^{6}\frac{1}{(2\ell _{i}+1)!!}\int_{0}^{\infty}dk\,k^{2+\ell _{1}+\ell _{2}}\,e^{-4\sqrt{k}}\int_{0}^{\infty}dk'\,k'^{\,2+\ell _{3}+\ell _{4}}\,e^{-4\sqrt{k'}}\int_{0}^{\infty}dk''\,k''^{\,2+\ell _{5}+\ell _{6}}\,e^{-4\sqrt{k''}}\nonumber\\
    &&\ \ \left[\frac{1}{(\sqrt{k}+\sqrt{k'})^{2}(\sqrt{k'}+\sqrt{k''})^{2}}\right]
    \Big[7k'+4k^{1/2}k'^{1/2}(6k'^{1/2}+k''^{1/2})\Big]\,.
\end{eqnarray}
On the other hand, the result of the integrations over the solid angles $\hat{k}$, $\hat{k}'$, and $\hat{k}''$ can be expressed in terms of another function,
\begin{eqnarray}\label{calE}
    {\cal E}_{\ell _{1}m_{1}\cdots \ell _{6}m_{6}}&\equiv& \int\,d\hat{k} d\hat{k'}d\hat{k''}\,Y_{\ell _{1}m_{1}}^{*}(\hat{k})Y_{\ell _{2}m_{2}}(\hat{k})Y_{\ell _{3}m_{3}}^{*}(\hat{k'})Y_{\ell _{4}m_{4}}(\hat{k'})Y_{\ell _{5}m_{5}}^{*}(\hat{k''})Y_{\ell _{6}m_{6}}(\hat{k''}) \nonumber\\
    &&\ \ (\hat{\epsilon}_{R}(\hat{k})\cdot\hat{\epsilon}_{R}(-\hat{k}'))
    (\hat{\epsilon}_{R}(\hat{k}')\cdot\hat{\epsilon}_{R}(-\hat{k}''))(\hat{\epsilon}_{R}(\hat{k}'')\cdot\hat{\epsilon}_{R}(-\hat{k}))\,.
\end{eqnarray}
The numerical values of ${\cal R}$ and ${\cal E}$ for various $\ell $ and $m$ have been tabulated in Appendix \ref{REvalues}.

With the help of ${\cal R}$ and ${\cal E}$, the connected 4PCF in Eq.~(\ref{4ptcf1}) can now be expressed as
\begin{eqnarray}\label{con4ptY}
&&\left\langle\phi(\tau,\vec{x})\phi(\tau,\vec{x}')\phi(\tau,\vec{x}'')\phi(\tau,\vec{x}''')\right\rangle_{conn}\nonumber\\
&=&\left(\frac{\alpha^{4}}{f^{4}}e^{8\pi\xi}\right)\left(\frac{15H^{8}}{2^{31}\pi^{5}\xi^{16}}\right)\nonumber\\
&&\ \ \sum_{\ell _{1}m_{1}\cdots \ell _{6}m_{6}}(-1)^{\ell _{1}+\ell _{3}+\ell _{5}}[(2\xi(-\tau)]^{-\sum_{i=1}^{6}\ell _{i}}\,
{\cal R}_{\ell _{1}\cdots \ell _{6}}{\cal E}_{\ell _{1}m_{1}\cdots \ell _{6}m_{6}}\,x^{\ell _{1}}x'^{\,\ell _{2}+\ell _{3}}x''^{\,\ell _{4}+\ell _{5}}x'''^{\,\ell _{6}}\nonumber\\
&&\hskip 80pt Y_{\ell _{1}m_{1}}(\hat{x})Y_{\ell _{2}m_{2}}^{*}(\hat{x}')Y_{\ell _{3}m_{3}}(\hat{x}')Y_{\ell _{4}m_{4}}^{*}(\hat{x}'')Y_{\ell _{5}m_{5}}(\hat{x}'')Y_{\ell _{6}m_{6}}^{*}(\hat{x}''')\nonumber\\
&&+\;23\ {\rm permutations\ of }\ \vec{x},\vec{x}',\vec{x}'',\vec{x}'''\,.
\end{eqnarray}

\section{Three-argument isotropic basis functions}


We expand 4PCFs in terms of the three-argument isotropic basis functions, which are 
\begin{eqnarray}\label{defcalP}
{\cal P}_{\ell_1\ell_2\ell_3}(\hat{x}_1,\hat{x}_2,\hat{x}_3) 
&=& \sum_{m_1m_2m_3} {\cal C}^{\ell_1\ell_2\ell_3}_{m_1m_2m_3} Y_{\ell_1m_1}(\hat{x}_1)Y_{\ell_2m_2}(\hat{x}_2)Y_{\ell_3m_3}(\hat{x}_3)\,,
\label{eqn:isofunc_4pcf}
\end{eqnarray}
where the weight is
\begin{eqnarray}
{\cal C}^{\ell_1\ell_2\ell_3}_{m_1m_2m_3} \equiv (-1)^{\ell_1+\ell_2+\ell_3}\six{\ell_1}{\ell_2}{\ell_3}{m_1}{m_2}{m_3}.
\label{eqn:clebsch_gorden_N3}
\end{eqnarray}
Here the Wigner-3j symbol is found in Appendix~\ref{sec:3j}. Under the parity operator, denoted by
$\hat\mathbb{P}$, the isotropic basis functions transform as
\begin{eqnarray}
\hat\mathbb{P}\left[{\cal P}_{\ell_1\ell_2\ell_3}(\hat{x}_1,\hat{x}_2,\hat{x}_3)\right] &\equiv& {\cal P}_{\ell_1\ell_2\ell_3}(-\hat{x}_1,-\hat{x}_2,-\hat{x}_3)\nonumber\\
&=& (-1)^{\ell_1+\ell_2+\ell_3}{\cal P}_{\ell_1\ell_2\ell_3}(\hat{x}_1,\hat{x}_2,\hat{x}_3)=
{\cal P}^*_{\ell_1\ell_2\ell_3}(\hat{x}_1,\hat{x}_2,\hat{x}_3).
\end{eqnarray}
Thus, when $\ell_1 + \ell_2 + \ell_3$ is even, they are real and even in parity. When $\ell_1 + \ell_2 + \ell_3$ is odd, they are pure imaginary and odd in parity.
Also, we have the orthogonal relation,
\begin{eqnarray}\label{isoortho}
\int d\hat{x}_1 d\hat{x}_2 d\hat{x}_3\, {\cal P}_{\ell_1\ell_2\ell_3}(\hat{x}_1,\hat{x}_2,\hat{x}_3)\; 
{\cal P}^*_{\ell'_1\ell'_2\ell'_3}(\hat{x}_1,\hat{x}_2,\hat{x}_3) = \delta_{\ell_1\ell'_1}\delta_{\ell_2\ell'_2}\delta_{\ell_3\ell'_3}.
\end{eqnarray}

Now we expand the 4PCF in Eq.~(\ref{con4ptY}) by even-parity and odd-parity isotropic basis functions as
\begin{eqnarray}
\zeta(\vec{x}',\vec{x}'',\vec{x}''')&\equiv& 
\langle \phi(\tau,\vec{0})\phi(\tau,\vec{x}')\phi(\tau,\vec{x}'')\phi(\tau,\vec{x}'''))\rangle \nonumber \\
&=& \sum_{\ell'\ell'' \ell'''} \zeta_{\ell'\ell''\ell'''}(x',x'',x''') 
{\cal P}_{\ell'\ell''\ell'''}(\hat{x}',\hat{x}'',\hat{x}''')\,,
\end{eqnarray}
where the expansion coefficients depend only on the $x_i$ and are given by orthogonality as
\begin{eqnarray}\label{zetalll}
\zeta_{\ell'\ell''\ell'''}(x',x'',x''') = \int d\hat{x}' d\hat{x}'' d\hat{x}'''\, 
\zeta(\vec{x}',\vec{x}'',\vec{x}''')\,
{\cal P}^*_{\ell'\ell''\ell'''}(\hat{x}',\hat{x}'',\hat{x}''').
\end{eqnarray}
To avoid an over-complete basis, the radial arguments $x$'s are ordered as $x'\leq x''\leq x'''$.
Since the 4PCF is real by definition, the expansion coefficients $\zeta_{\ell'\ell''\ell'''}(x',x'',x''')$ are real when $\ell' + \ell'' + \ell'''$ is even and pure imaginary when the sum is odd.

Therefore, non-zero expansion coefficients $\zeta_{\ell'\ell''\ell'''}$ with odd $\ell'+\ell''+\ell'''$ will indicate the presence of parity-violating effects. These coefficients can be calculated theoretically by decomposing the 4PCF in terms of the isotropic basis functions as we have mentioned above for any particular model. Here, in this paper we have concentrated on the axion inflation one. On the other hand, from the data of observational measurements on the distributions of galaxies, one can also derive these coefficients. Non-zero coefficients with odd $\ell'+\ell''+\ell'''$ not only indicate parity violation, but their magnitudes can in turn constrain the parameters in the particular model under consideration. This we shall do in the next section.


To obtain these coefficients, we insert the expression of the 4PCF in Eq.~(\ref{con4ptY}) into Eq.~(\ref{zetalll}). The integrations of the spherical harmonics over the solid angles $\hat{x}'$, $\hat{x}''$, and $\hat{x}'''$ can be evaluated using the formulas in Eqs.~(\ref{YYint}) and (\ref{YYYint}). Then, the expansion coefficient becomes
\begin{eqnarray}\label{zeta4pt}
    &&\zeta_{\ell'\ell''\ell'''}(x',x'',x''')\nonumber\\
    &=&\left(\frac{\alpha^{4}}{f^{4}}e^{8\pi\xi}\right)\left(\frac{15H^{8}}{2^{32}\pi^{11/2}\xi^{16}}\right)\nonumber\\
&&\ \ \sum_{\ell _{2}m_{2}\cdots \ell _{5}m_{5}}\sum_{m'm''m'''}i^{\ell_{2}+\cdots+\ell_{5}+\ell'''}(-1)^{\ell _{3}+\ell _{5}+\ell'+\ell''+\ell'''+m_{2}+m_{4}+m'+m''+m'''}\,[(2\xi(-\tau)]^{-\ell_{2}-\cdots-\ell_{5}-\ell'''}\nonumber\\
&&\hskip 100pt \sqrt{\frac{(2\ell_2+1)(2\ell_3+1)(2\ell'+1)}{4\pi}}\sqrt{\frac{(2\ell_4+1)(2\ell_5+1)(2\ell''+1)}{4\pi}}\nonumber\\
&&\hskip 10pt
\six{\ell'}{\ell''}{\ell'''}{m'}{m''}{m'''}
\six{\ell_2}{\ell_3}{\ell'}{0}{0}{0}
\six{\ell_2}{\ell_3}{\ell'}{-m_2}{m_3}{-m'}
\six{\ell_4}{\ell_5}{\ell''}{0}{0}{0}
\six{\ell_4}{\ell_5}{\ell''}{-m_4}{m_5}{-m''}
\nonumber\\
&&\hskip 100pt x'^{\,\ell _{2}+\ell _{3}}x''^{\,\ell _{4}+\ell _{5}}x'''^{\,\ell'''}{\cal R}_{0\ell_2\cdots \ell _{5}\ell'''}{\cal E}_{00\ell _{2}m_{2}\cdots \ell _{5}m_{5}\ell'''\,-m'''}\nonumber\\
&&+\ 23\ {\rm permutation\ terms}\,.\ 
\end{eqnarray}
These coefficients can be further simplified by considering the selection rules of the Wigner-3j symbols as mentioned in Appendix~\ref{sec:3j}. In particular, $\six{\ell_1}{\ell_2}{\ell_3}{m_1}{m_2}{m_3}$ is nonzero only if $|\ell_1-\ell_2|\leq \ell_3\leq\ell_1+\ell_2$ and $m_1+m_2+m_3=0$. Applying these rules to the third and the fifth Wigner-3j symbols in Eq.~(\ref{zeta4pt}), we see that $|\ell_2-\ell_3|\leq \ell'\leq\ell_2+\ell_3$ and $|\ell_4-\ell_5|\leq \ell''\leq\ell_4+\ell_5$. We note that we have a factor $\xi^{-\ell_2-\cdots\ell_4-\ell'''}$. In the large $\xi$ approximation, we like to have the power of $\xi$ to be as small as possible. Hence, from the selection rules, we need to take $\ell_2+\ell_3=\ell'$ and $\ell_3+\ell_4=\ell''$. Also, from the selection rules on $m$ for these two Wagner-3j symbols, we have $m_3=m_2+m'$ and $m_5=m_4+m''$. Taking these into consideration, we have
\begin{eqnarray}\label{term0123}
    &&\zeta_{\ell'\ell''\ell'''}(x',x'',x''')\nonumber\\
    &=&\left(\frac{\alpha^{4}}{f^{4}}e^{8\pi\xi}\right)\left(\frac{15H^{8}}{2^{34}\pi^{13/2}\xi^{16}}\right)i^{\ell'+\ell''+\ell'''}\left[2\xi(-\tau)\right]^{-\ell'-\ell''-\ell'''}x'^{\ell'}x''^{\ell''}x'''^{\ell'''}
    \nonumber\\
&&\ \ 
\sum_{m'm''m'''}\sum_{\ell _{2}m_{2}\ell _{4}m_{4}}(-1)^{\ell _{2}+\ell _{4}+\ell'''+m_{2}+m_{4}+m'+m''+m'''}\nonumber\\
&&\hskip 60pt \sqrt{(2\ell_2+1)(2(\ell'-\ell_2)+1)(2\ell'+1)}\sqrt{(2\ell_4+1)(2(\ell''-\ell_4)+1)(2\ell''+1)}\nonumber\\
&&\hskip 60pt
\six{\ell'}{\ell''}{\ell'''}{m'}{m''}{m'''}
\six{\ell_2}{\ell'-\ell_2}{\ell'}{0}{0}{0}
\six{\ell_2}{\ell'-\ell_2}{\ell'}{-m_2}{m_2+m'}{-m'}\nonumber\\
&&\hskip 60pt
\six{\ell_4}{\ell''-\ell_4}{\ell''}{0}{0}{0}
\six{\ell_4}{\ell''-\ell_4}{\ell''}{-m_4}{m_4+m''}{-m''}
\nonumber\\
&&\hskip 60pt 
{\cal R}_{0\,\ell_2\,\ell'-\ell_{2}\,\ell_3\,\ell''-\ell_3\,\ell'''}\,{\cal E}_{00\,\ell _{2}m_{2}\,\ell'-\ell _{2}m_{2}+m'\,\ell_4 m_4\, \ell''-\ell_4\,m_4+m''\,\ell'''-m'''}\nonumber\\
&&+\ 23\ {\rm permutation\ terms}\,.\ 
\end{eqnarray}

The other 23 permutation terms can be evaluated in a similar fashion. In total, there are 12 terms with 0 for $\ell_1$ or $\ell_6$. The expression for these terms is like the one shown in Eq.~(\ref{term0123}) with 5 Wigner-3j symbols. The rest 12 terms with 0 for $(\ell_2,\ell_3)$ or $(\ell_4,\ell_5)$ would have only 3 Wigner-3j symbols. Now, for the 6 terms with 0 for $\ell_1$, they can be obtained by permuting $\ell'$, $\ell''$, and $\ell'''$ for the term shown in Eq.~(\ref{term0123}). The same consideration can be applied to the 6 terms with 0 for $\ell_6$, 0 for $(\ell_2,\ell_3)$, 0 for $(\ell_4,\ell_5)$.
Putting them together, we have the final expression for the expansion coefficient,
\begin{eqnarray}\label{finalzeta}
     &&\zeta_{\ell'\ell''\ell'''}(x',x'',x''')\nonumber\\
    &=&\left(\frac{\alpha^{4}}{f^{4}}e^{8\pi\xi}\right)\left(\frac{15H^{8}}{2^{34}\pi^{13/2}\xi^{16}}\right)i^{\ell'+\ell''+\ell'''}\left[2\xi(-\tau)\right]^{-\ell'-\ell''-\ell'''}x'^{\ell'}x''^{\ell''}x'''^{\ell'''}
    \nonumber\\
&&\sum_{m'm''m'''}\six{\ell'}{\ell''}{\ell'''}{m'}{m''}{m'''}\Big({\cal Z}_{\ell'\, m'\ell'' \,m'' \ell''' \,m'''}
+\ 5\ {\rm permutations\ of}\ \ell'\,m', \ell'' m''\ {\rm and}\ \ell''' m'''\Big),\nonumber\\
\end{eqnarray}
where
\begin{eqnarray}
     &&{\cal Z}_{\ell'\, m'\ell'' \,m'' \ell''' \,m'''}\nonumber\\
&&\sum_{\ell _{1}m_{1}}(-1)^{\ell _{1}+m_{1}}\sqrt{(2\ell_1+1)(2(\ell'-\ell_1)+1)(2\ell'+1)}\nonumber\\
&&\hskip 60pt\six{\ell_1}{\ell'-\ell_1}{\ell'}{0}{0}{0}
\six{\ell_1}{\ell'-\ell_1}{\ell'}{-m_1}{m_1+m'}{-m'}
\nonumber\\
&&\ \ \Bigg[\ (-1)^{\ell'''+m'+m'''}{\cal R}_{\ell''\,0\,0\,\ell_1\,\ell'-\ell_{1}\,\ell'''}\,{\cal E}_{\ell'' m''\,00\,00\,\ell _{1}m_{1}\,\ell'-\ell _{1}m_{1}+m'\,\ell''' -m'''}\nonumber\\
&&\ \ \ \ +(-1)^{\ell'''+m'+m'''}{\cal R}_{\ell''\,\ell_1\,\ell'-\ell_{1}\,0\,0\,\ell'''}\,{\cal E}_{\ell'' m''\,\ell _{1}m_{1}\,\ell'-\ell _{1}m_{1}+m'\,00\,00\,\ell''' -m'''}\nonumber\\
&&\ \ \ \ +\sum_{\ell _{2}m_{2}}\sqrt{(2\ell_2+1)(2(\ell''-\ell_2)+1)(2\ell''+1)}\nonumber\\
&&\hskip 60pt\six{\ell_2}{\ell''-\ell_2}{\ell''}{0}{0}{0}
\six{\ell_2}{\ell''-\ell_2}{\ell''}{-m_2}{m_2+m''}{-m''}
\nonumber\\
&&\ \ \ \ \ \Bigg((-1)^{\ell _{2}+\ell'''+m_{2}+m'+m''+m'''}{\cal R}_{0\,\ell_1\,\ell'-\ell_{1}\,\ell_2\,\ell''-\ell_2\,\ell'''}\,{\cal E}_{00\,\ell _{1}m_{1}\,\ell'-\ell _{1}m_{1}+m'\,\ell_2 m_2\, \ell''-\ell_2\,m_2+m''\,\ell'''-m'''}\nonumber\\
&&\hskip 30pt 
+(-1)^{\ell _{2}+m_{2}+m'+m''}{\cal R}_{\ell'''\,\ell_1\,\ell'-\ell_{1}\,\ell_2\,\ell''-\ell_2\,0}\,{\cal E}_{\ell''' m'''\,\ell _{1}m_{1}\,\ell'-\ell _{1}m_{1}+m'\,\ell_2 m_2\, \ell''-\ell_2\,m_2+m''\,00}\Bigg)\Bigg].\nonumber\\
\end{eqnarray}
As we have discussed above, $\zeta_{\ell',\ell'',\ell'''}(x',x'',x''')$ are real when $\ell'+\ell''+\ell'''$ is even and pure imaginary when the sum is odd. It is apparent for the leading $\xi$ expression we have in Eq.~(\ref{finalzeta}) that this is the case as it is proportional to $i^{\ell'+\ell''+\ell'''}$ and the rest of the expression is real.

With the expression in Eq.~(\ref{finalzeta}), we can work out the leading $\xi$ behavior of the expansion coefficient $\zeta_{\ell',\ell'',\ell'''}(x',x'',x''')$ for specific values of $\ell'$, $\ell''$, and $\ell'''$. First, for $\ell'=\ell''=\ell'''=0$, we have
\begin{eqnarray}
    \zeta_{000}(x',x'',x''')&=&\left(\frac{\alpha^{4}}{f^{4}}e^{8\pi\xi}\right)\left(\frac{15H^{8}}{2^{34}\pi^{13/2}\xi^{16}}\right)(6)(4\,{\cal R}_{000000}\,{\cal E}_{00\cdots 00})\nonumber\\
    &=&\left(\frac{\alpha^{4}}{f^{4}}e^{8\pi\xi}\right)\left(\frac{15H^{8}}{2^{34}\pi^{13/2}\xi^{16}}\right)\left(\frac{8}{3}\,{\cal R}_{000000}\right)\nonumber\\
    &\approx&\left(\frac{\alpha^{4}}{f^{4}}e^{8\pi\xi}\right)\left(\frac{15H^{8}}{2^{34}\pi^{13/2}\xi^{16}}\right)(0.000755235),
\end{eqnarray}
where we have used Eqs.~(\ref{coeffinal}) and (\ref{epsiloncoef}) to get
\begin{eqnarray}
    {\cal E}_{00\cdots 00}=\epsilon_{000}\left(\frac{1}{8\pi^{3/2}}\right)=\frac{1}{9}
\end{eqnarray}
and the explicit value of ${\cal R}_{000000}$ is given in Eq.~(\ref{R0}) or Table~\ref{Rtable}.

Since we are interested in the possibility of parity violation effects manifesting in the 4PCF, we shall thus focus our attention to the cases with odd $\ell'+\ell''+\ell'''$. Moreover, we shall evaluate the expansion coefficients $\zeta_{\ell'\ell''\ell'''}$ with $\ell'\leq\ell''\leq\ell'''$. The other coefficients can be obtained by considering the symmetry inherited from the definitions of ${\cal P}_{\ell_1\ell_2\ell_3}(\hat{x}_1,\hat{x}_2,\hat{x}_3)$ and $\zeta_{\ell'\ell''\ell'''}(x',x'',x''')$ in Eqs.~(\ref{defcalP}) and (\ref{zetalll}), respectively.

\subsection{$\zeta_{111}$}
To start with, we look at $\zeta_{\ell'\ell''\ell'''}$ with $\ell'=\ell''=\ell'''=1$. From Eq.~(\ref{finalzeta}), we note that the Wigner-3j symbols are nonzero only for the six sets of $(m',m'',m''')=(1,0,-1),\,(0,-1,1),\,(-1,1,0),\,(1,-1,0),\,(-1,0,1),\,(1,0,-1)$. If we concentrate on the expression for $m'=1,m''=0,m'''=-1$, we have
\begin{eqnarray}
    &&\six{1}{1}{1}{1}{0}{-1}\Big[{\cal Z}_{11\,10\,1\,-\!1}
+\ 5\ {\rm permutations\ of}\ (11), (10)\ {\rm and}\ (1\ -\!1)\Big]=0.
\end{eqnarray}
It is easy to verify that the results for the other five sets of $(m',m'',m''')$ are all vanishing. Therefore, we have the final result that
\begin{eqnarray}
    \zeta_{111}(x',x'',x''')=0.
\end{eqnarray}

\subsection{$\zeta_{122}$}
In this case, we have $m'=(1,0,-1)$ and $m'',m'''=(2,1,0,-1,-2)$. Again due to the symmetry of the Wigner-3j symbols, the only sets of $(m',m'',m''')$ which are non-vanishing are $(1,1,-2),\,(1,-2,1),\,(1,0,-1),\,(1,-1,0),\,(0,2,-2),\,(0,-2,2),\,(0,1,-1),\,(0,-1,1)$, $(-1,2,-1),\,(-1,-1,2),\,(-1,1,0),\,(-1,0,1)$. For $(1,1,-2)$, we have the expression in Eq.~(\ref{finalzeta}) as
\begin{eqnarray}\label{term11-2}
     &&\six{1}{2}{2}{1}{1}{-2}\Big[{\cal Z}_{11\,21\,2\,-\!2}
+\ 5\ {\rm permutations\ of}\ (11), (21)\ {\rm and}\ (2\ -\!2)\Big]\nonumber\\
&=&\frac{1}{9000}\,\bigg[2\sqrt{10}\,\big(\,5\,{\cal R}_{102020}-3\,{\cal R}_{111020}-5\,{\cal R}_{201020}\big)\nonumber\\
&&\hskip 50pt+\sqrt{3}\,(1+\sqrt{2})\big(3\,{\cal R}_{111110}-25\,{\cal R}_{102110}+5\,{\cal R}_{201110}\big)\bigg]
\end{eqnarray}
Consider now the set $(1,-2,1)$. Here, we have
\begin{eqnarray}\label{term1-21}
    &&\six{1}{2}{2}{1}{-2}{1}\Big[{\cal Z}_{11\,2\,-\!2\ 21}
+\ 5\ {\rm permutations\ of}\ (11), (2\ -\!2)\ {\rm and}\ (21)\Big].
\end{eqnarray}
From the symmetry of the Wigner-3j symbols, 
\begin{eqnarray}
    \six{1}{2}{2}{1}{-2}{1}=-\six{1}{2}{2}{1}{1}{-2}=\frac{1}{\sqrt{15}},
\end{eqnarray}
while due to the permutations the expressions in the square brackets in Eqs.~(\ref{term11-2}) and (\ref{term1-21}) are actually the same. Hence, the two expressions in Eqs.~(\ref{term11-2}) and (\ref{term1-21}) cancel each other. In the same way, the other sets of $(m',m'',m''')$ will cancel pairwise. For example, the expressions corresponding to $(1,0,-1)$ and $(1,-1,0)$ will differ by a minus sign and add up to zero. We thus have
\begin{eqnarray}
    \zeta_{122}(x',x'',x''')=0.
\end{eqnarray}

Note that what we have been considering is the leading $1/\xi$ order contribution to the expansion coefficient. In this order, the coefficient $\zeta_{\ell'\ell''\ell'''}$ is proportional to $(1/\xi)^{16+\ell'+\ell''+\ell'''}$ as evident from Eq.~(\ref{finalzeta}). However, from the previous results on $\zeta_{111}$ and $\zeta_{122}$, it is quite apparent that the parity-violating expansion coefficients, that is, with odd $\ell'+\ell''+\ell'''$, are zero in this order as long as they have repeated indices. Therefore, in the following we shall work on $\zeta_{234}$ which has no repeated indices and is expected to be nonzero.

\subsection{$\zeta_{234}$}
For $\zeta_{234}$, due to the symmetry of the Wigner-3j symbols, there are all together 32 sets of $(m',m'',m''')$ with nonzero value including $(2,2,-4),(2,1,-3)$, etc. First, for $(2,2,-4)$, we have from Eq.~(\ref{finalzeta})
\begin{eqnarray}
     &&\six{2}{3}{4}{2}{2}{-4}\Big[{\cal Z}_{22\,32\,4\,-\!4}
+\ 5\ {\rm permutations\ of}\ (22), (32)\ {\rm and}\ (4\ -\!4)\Big]\nonumber\\
&=&\frac{1}{1050}{\cal R}_{011124}+\frac{1}{1350}{\cal R}_{011133}-\frac{1}{210}{\cal R}_{011223}+\frac{1}{525}{\cal R}_{011313}-\frac{1}{1029}{\cal R}_{012222}+\frac{1}{210}{\cal R}_{012312}\nonumber\\
&&-\frac{1}{630}{\cal R}_{012402}-\frac{1}{810}{\cal R}_{013302}-\frac{1}{630}{\cal R}_{020124}-\frac{1}{810}{\cal R}_{020133}+\frac{1}{630}{\cal R}_{020223}+\frac{1}{735}{\cal R}_{021222}\nonumber\\
&&-\frac{1}{525}{\cal R}_{021312}+\frac{1}{630}{\cal R}_{022302}-\frac{1}{630}{\cal R}_{200124}-\frac{1}{810}{\cal R}_{200133}+\frac{1}{630}{\cal R}_{200223}+\frac{\sqrt{70}}{6300}{\cal R}_{203220}\nonumber\\
&&-\frac{\sqrt{3}}{1620}{\cal R}_{203310}-\frac{\sqrt{21}}{1890}{\cal R}_{204210}-\frac{\sqrt{42}}{3150}{\cal R}_{213120}+\frac{\sqrt{21}}{630}{\cal R}_{213210}+\frac{\sqrt{42}}{4410}{\cal R}_{222120}\nonumber\\
&&-\frac{\sqrt{42}}{6174}{\cal R}_{222210}-\frac{\sqrt{21}\,(2-\sqrt{2})}{6300}{\cal R}_{231120}-\frac{\sqrt{21}\,(2-\sqrt{2})}{1260}{\cal R}_{231210}+\frac{\sqrt{21}\,(2-\sqrt{2})}{3780}{\cal R}_{240120}\nonumber\\
&&+\frac{\sqrt{21}\,(2-\sqrt{2})}{3780}{\cal R}_{240210}+\frac{\sqrt{30}}{3150}{\cal R}_{313110}+\frac{1}{630}{\cal R}_{322002}+\frac{1}{630}{\cal R}_{322020}-\frac{\sqrt{30}}{1260}{\cal R}_{322110}\nonumber\\
&&-\frac{1}{810}{\cal R}_{331002}-\frac{1}{810}{\cal R}_{331020}+\frac{\sqrt{30}}{8100}{\cal R}_{331110}+\frac{\sqrt{21}\,(2-\sqrt{2})}{3780}{\cal R}_{420120}\nonumber\\
&&+\frac{\sqrt{21}\,(2-\sqrt{2})}{3780}{\cal R}_{420210}-\frac{1}{630}{\cal R}_{421002}-\frac{1}{630}{\cal R}_{421020}+\frac{\sqrt{30}}{6300}{\cal R}_{421110}\,.
\end{eqnarray}
We also find that the result from $(-2,-2,4)$ is the same as that from $(2,2,-4)$ above. The other 30 sets of $(m',m'',m''')$ can be evaluated in a similar way.

Putting all the results together, we have finally
\begin{eqnarray}\label{zeta234}
    &&\zeta_{234}(x',x'',x''')\nonumber\\
    &=&\left(\frac{\alpha^{4}}{f^{4}}e^{8\pi\xi}\right)\left(\frac{15H^{8}}{2^{34}\pi^{13/2}\xi^{16}}\right)(i)\left[2\xi(-\tau)\right]^{-9}(x')^{2}(x'')^{3}(x''')^{4}
    \nonumber\\
    &&\Bigg[\frac{3}{140}{\cal R}_{011124}+\frac{1}{60}{\cal R}_{011133}-\frac{3}{28}{\cal R}_{011223}+\frac{3}{70}{\cal R}_{011313}-\frac{15}{686}{\cal R}_{012222}+\frac{3}{28}{\cal R}_{012312}\nonumber\\
&&-\frac{1}{28}{\cal R}_{012402}-\frac{1}{36}{\cal R}_{013302}-\frac{1}{28}{\cal R}_{020124}-\frac{1}{36}{\cal R}_{020133}+\frac{1}{28}{\cal R}_{020223}+\frac{3}{98}{\cal R}_{021222}\nonumber\\
&&-\frac{3}{70}{\cal R}_{021312}+\frac{1}{28}{\cal R}_{022302}-\frac{1}{28}{\cal R}_{200124}-\frac{1}{36}{\cal R}_{200133}+\frac{1}{28}{\cal R}_{200223}\nonumber\\
&&+\frac{(8-2\sqrt{5}+2\sqrt{10}+2\sqrt{15}-2\sqrt{30}-2\sqrt{35}+\sqrt{70})}{1260}{\cal R}_{203220}\nonumber\\
&&+\frac{(21-7\sqrt{3}-3\sqrt{7}-3\sqrt{14}-3\sqrt{21}+\sqrt{210}}{4536}{\cal R}_{203310}\nonumber\\
&&+\sqrt{7}(30-10\sqrt{3}-6\sqrt{5}-10\sqrt{6}+6\sqrt{10}-\sqrt{15}+4\sqrt{30})\nonumber\\
&&\ \ \left(\frac{1}{8820}{\cal R}_{204210}+\frac{1}{7350}{\cal R}_{213120}-\frac{1}{2940}{\cal R}_{213210}-\frac{1}{10290}{\cal R}_{222120}+\frac{1}{14406}{\cal R}_{222210}\right)\nonumber\\
&&+\frac{1}{28}{\cal R}_{322002}+\frac{1}{140}{\cal R}_{322020}-\frac{1}{36}{\cal R}_{331002}-\frac{1}{180}{\cal R}_{331020}-\frac{1}{28}{\cal R}_{421002}-\frac{1}{140}{\cal R}_{421020}\nonumber\\
&&+\sqrt{5}(2-2\sqrt{3}+\sqrt{6})\left(\frac{1}{350}{\cal R}_{313110}-\frac{1}{140}{\cal R}_{322110}+\frac{1}{900}{\cal R}_{331110}+\frac{1}{700}{\cal R}_{421110}\right)\Bigg]\nonumber\\
&\approx&\left(\frac{\alpha^{4}}{f^{4}}e^{8\pi\xi}\right)\left(\frac{15H^{8}}{2^{34}\pi^{13/2}\xi^{16}}\right)(i)\left[2\xi(-\tau)\right]^{-9}(x')^{2}(x'')^{3}(x''')^{4}(-0.000148496).
\end{eqnarray}
If we take the ratio between $\zeta_{234}$ and $\zeta_{000}$,
\begin{eqnarray}\label{zetaratio}
    \frac{\zeta_{234}(x',x'',x''')}{\zeta_{000}(x',x'',x''')}
    &\approx&(i)\left[2\xi(-\tau)\right]^{-9}(x')^{2}(x'')^{3}(x''')^{4}\left(\frac{-0.000148496}{0.000755235}\right)\nonumber\\
    &\approx&\left(-\frac{i(x')^2(x'')^3(x''')^4}{\xi^9(-\tau)^9}\right)(0.000384028). 
\end{eqnarray}

Finally, we obtain a non-zero coefficient with odd $\ell'+\ell''+\ell'''$ indicating that there are parity-violating effects in the axion inflation model that we are considering. The other coefficients with even or odd $\ell'+\ell''+\ell'''$ can be calculated in the large $\xi$ approximation along the same lines as above. However, it is quite apparent that the procedure becomes more and more tedious as the numbers $\ell'$, $\ell''$, and $\ell'''$ increase. 

\section{Observational constraints on parity violation}

Recently, there have been claims about the detection of the non-Gaussian 4PCF of the galaxy distribution using 
the Baryon Oscillation Spectroscopic Survey (BOSS) data of Sloan Digital Sky Survey (SDSS) \cite{PHS21,HSC23,Philcox22}. 
Parity-even modes in the 4PCF are detected at $8.1\sigma$ \cite{PHS21}. In LOWZ galaxies, they find $3.1\sigma$ evidence for a non-zero parity-odd 4PCF, and in CMASS galaxies they detect a parity-odd 4PCF at $7.1\sigma$ \cite{HSC23} and at $2.9\sigma$ \cite{Philcox22}. This provides significant evidence for cosmological parity violation. 
However, Ref.~\cite{PE25} has reported that the parity-violating signal seen in the BOSS data may suffer from large systematic uncertainties.

To make a comparison with the galaxy observation, we need to link the inflaton fluctuations to the galaxy number counts. As known in slow-roll inflation, the primordial density power spectrum is given by $\delta\rho(k)= (H/{\dot\Phi}) \phi(k)$. The fluctuations in the galaxy number counts at the present time can be originated from the primordial density power spectrum by
the relation, $\delta n(k)=b(k) T(k) \delta\rho(k)$, 
where $b(k)$ is the bias factor and $T(k)$ the transfer function.
Since we are interested in the ratios of the expansion coefficients at a particular length scale, $H/{\dot\Phi}$, $b(k)$, and $T(k)$ are factored out and thus we can directly use the measurements of the galaxy number counts.

In the previous section, we have evaluated with great effort some low-order expansion coefficients. Interestingly, this suffices to have a simple test on the axion inflation model that saliently predicts a parity-violating 4PCF. 
For the parity-odd modes, Refs.~\cite{HSC23,Philcox22} have shown measurements of low-order expansion coefficients with repeated indices and the one with $\zeta_{234}$ (in Ref.~\cite{Philcox22}) in a series of radial bins, representing various galaxy separations in the range $[20,160]\,h^{-1} {\rm Mpc}$. Apparently, the expansion coefficients with repeated indices are mostly consistent with zero within observational uncertainties, except that certain radial bins contain significant parity-violating signals. These signals seem to rule out the axion inflation model being responsible for the main generating source, as we have shown in the previous section that the expansion coefficients vanish as long as they have repeated indices. 
However, we anticipate that coefficients like $\zeta_{111}$ or $\zeta_{122}$, which vanish in the leading $1/\xi$ approximation, will have non-zero values at higher orders, likely because cancellations observed in the leading order may not occur. Also, coefficients like $\zeta_{234}$, which are already non-zero in the leading order, should receive higher-order contributions. In the following, we will compare the results obtained in the previous section with the observational data, making an attempt to derive constraints on the axion inflation model.

First of all, the zero values of $\zeta_{111}$ and $\zeta_{122}$ are consistent with the measurements within the observational uncertainties in most radial bins, as shown in Figs.~5 and~6 of Ref.~\cite{HSC23} and Fig.~2 of Ref.~\cite{Philcox22}. In Fig.~2 of Ref.~\cite{Philcox22}, the measurement of $\zeta_{234}$ is consistent with zero and within $-200<x_1 x_2 x_3 {\rm Im}\,\zeta_{234}<200$. For the parity-even $x_1 x_2 x_3\zeta_{000}$, the first five radial bins in Fig.~3 of Ref.~\cite{PHS21} show that $x_1 x_2 x_3\zeta_{000} \sim 1000$ for galaxy separations in the range $[20,80]\,h^{-1} {\rm Mpc}$. This gives a constraint on the ratio, 
\begin{eqnarray} \label{zeta_constraint}
 -0.2<\frac{{\rm Im}\,\zeta_{234}(x_1,x_2,x_3)}{\zeta_{000}(x_1,x_2,x_3)}<0.2 \quad{\rm for}\quad 20<\frac{x_i}{h^{-1} {\rm Mpc}}<80.
\end{eqnarray}

Defining $\rho\equiv-\xi\tau/x$, we have the model prediction in Eq.~(\ref{zetaratio}) as
\begin{eqnarray}
    \frac{{\rm Im}\,\zeta_{234}(x_1,x_2,x_3)}{\zeta_{000}(x_1,x_2,x_3)}\approx 
    -\frac{0.000384028}{\rho_1^{2} \rho_2^{3} \rho_3^{4}}\,.
\end{eqnarray}
Before taking $\tau\rightarrow 0$ at the end of inflation to find the value of this ratio, 
the reader should be reminded that the expression in Eq.~(\ref{Amode}) is an approximated solution to the mode function 
$A_{R}(k,\tau)$ in the range,
\begin{eqnarray}\label{xi_range}
\frac{1}{8\xi} \leq -k\tau \leq 2\xi, \quad{\rm or\; equivalently}\quad
\frac{1}{8\xi} \leq -\frac{2\pi\tau}{x} \leq 2\xi,
\end{eqnarray}
which corresponds to the period during which the right-handed gauge field is maximally amplified~\cite{BP11,AS12}. 
In order for the approximation to work, the value of $\rho$ must be at least $1/(16\pi)\simeq 0.02$. On the other hand, the constraint~(\ref{zeta_constraint}) implies that 
\begin{eqnarray}\label{rho_lowerbound}
\rho_1\sim \rho_2\sim \rho_3\gtrsim 0.5,
\end{eqnarray}
which lies within the range~(\ref{xi_range}). This shows that the result we have obtained for the expansion coefficient $\zeta_{234}$, within the allowed ranges of parameters in the axion inflation model, is consistent with the values derived from observed distributions of galaxies in various data samples.

The photon production in axion inflation is rather model-dependent, and the produced photons can act as a source for density perturbation, in addition to the de Sitter vacuum fluctuations. The production process can take place for a brief period of time during inflation, depending on the form of the axion potential, provided that induced large-scale density perturbation is constrained by cosmic microwave background data and that small-scale density perturbation would not overproduce primordial black holes. See, for example, Ref.~\cite{CLN18}, which shows that tuned modulations to the axion cosine potential can produce photons at desired timing during inflation, thus seeding the formation of primordial black holes with masses ranging from $10^8$ grams to $20$ solar masses. Let us assume a standard axion inflation that lasts about $60$ e-folds. Then, $a(\tau)=-1/H\tau=e^N$, with $N=0$ and $N=60$ denoting the onset and end of inflation, respectively. Thus, the length scales of $[20,80]\,h^{-1} {\rm Mpc}$ exit the horizon at about $N=7$. Inducing density perturbation of these length scales by the photon production should take place at the time that satisfies
\begin{eqnarray}
0.5 \lesssim \rho=-\frac{\xi\tau}{x}=\frac{\xi}{aHx}=
\frac{\xi\,e^7}{2\pi e^N}\leq \frac{\xi^2}{\pi}
\quad\Rightarrow\quad 7-\ln(2\xi)\leq N\lesssim 7+\ln\frac{\xi}{\pi}\,,
\end{eqnarray}
where we have substituted $x=2\pi e^{-7} H^{-1}$, and used the range~(\ref{xi_range}) and the lower bound~(\ref{rho_lowerbound}).
For example, taking $\xi=15$, we obtain $4.0\leq N\lesssim 8.6$.
In particular, the parity-odd modes are generated maximally as photon production occurs near $N=8.6$. 
In Appendix~\ref{sec:example}, we have provided a toy inflation model that gives rise to a narrow peak of density perturbation with $\xi\sim 15$ at $N\sim 9$.
It would be interesting to construct a realistic model in the axion monodromy inflation model~\cite{CLN18}.

\section{Conclusions and discussions}

We have investigated the possibility of parity violation in the early universe by studying the 4PCF within the axion inflation model. Our approach builds on our previous work \cite{ChoNg20}, which uses an open quantum system formalism to systematically compute inflaton in-in correlation functions to higher orders. In this formalism, photon degrees of freedom are integrated out to produce the influence functional. To obtain the 4PCF, we needed to calculate the noise kernel in this functional up to the fourth order. These calculations were organized by grouping the 48 connected terms into three types, based on how the various points of the photon Green functions are connected. The expressions simplify significantly in the large $\xi$ approximation, where a large $\xi$ enhances right-handed photons in the model.

Using isotropic basis functions \cite{CS23}, we decomposed the 4PCF, resulting in expansion coefficients $\zeta_{\ell'\ell''\ell'''}$ that naturally split into components even or odd under parity. Specifically, coefficients with an odd sum of indices $(\ell'+\ell''+\ell''')$ are parity-odd, and their non-zero values would indicate parity violation. In the leading $1/\xi$ approximation we worked with, we found that parity-odd coefficients with repeated indices, such as $\zeta_{111}$ and $\zeta_{122}$, vanish. On the other hand, we have obtained a non-zero $\zeta_{234}$, which provides a definite indication of parity violation in the axion inflation model. We compared these theoretical values for the $\zeta$ coefficients with those derived from recent galaxy survey data \cite{PHS21,HSC23,Philcox22}, 
and we obtained interesting constraints on the axion inflation model.

Due to the complexity of the multiple integrals involved in deriving the 4PCF, previous studies \cite{NRSX23,FMOS24,RSHG24,BWXZ25} of this problem have primarily relied on numerical techniques. In contrast, we developed a fully analytical approach. Using the $1/\xi$ approximation, both the photon Green functions and various radial integrals could be simplified. Additionally, we utilized isotropic basis functions to express the product of photon polarization vectors in a form that facilitated the evaluation of angular integrals using Wigner-3j symbols. This analytical method allows for much easier exploration of the parameter space in the model. As demonstrated in the last section, this enables the development of constraints on the strength of the axion-photon coupling and the timing of photon production during inflation.

As an immediate extension of our current work, this method can certainly be used to calculate other parity-odd coefficients, such as $\zeta_{245}$ and $\zeta_{256}$. The number of terms involved in the final summations for these higher $\zeta$ coefficients is expected to increase significantly compared to those in $\zeta_{234}$ (Eq. (59)). Another extension would involve working on higher orders in the $1/\xi$ approximation. This can be achieved by developing $1/\xi$ expansions for the Green functions and various integrands, especially those for radial integrations. 
We anticipate that coefficients with repeated indices, which vanish in the leading $1/\xi$ approximation, will have non-zero values at higher orders. Furthermore, this expansion could yield more accurate values for non-zero coefficients like $\zeta_{234}$.
Indeed, in axion inflation model, the value of $\xi$ is typically larger than $1$, so that photons can be copiously produced and induce a large density perturbation. However, the strong backreaction of photon production could slow down the inflaton speed, thus making $\xi$ much less than $1$ at some moments, while the density perturbation remains large~\cite{CLN16,NT16,DGWW20,PS23}.  
Therefore, it warrants a full analysis of a specific axion inflation model in order to set 
further, more precise constraints on it.
All of these efforts will be worthwhile when more precise measurements~\cite{CSH23} or new methods~\cite{GSHK25} become available, particularly from data showing definite parity-violating signals in the 4PCF. Importantly, it would allow for discriminating the axion model from other inflationary models~\cite{Shiraishi16,LTWX20,CJPS23,CIP23,CAKP23,STZ25,OAL25} and post-inflationary scenarios~\cite{PCM24,IJK25,YSSY25}.

\begin{acknowledgments}
The authors are thankful to Shu-Lin Cheng for the numerical work in Appendix~\ref{sec:example}.
This work was supported in part by the National Science and
Technology Council (NSTC) of Taiwan, Republic of
China, under Grant Nos. MOST 113-2112-M-032-007 (H.T.C.) 
and NSTC 113-2112-M-001-033 (K.W.N.).
\end{acknowledgments}

\appendix

\section{Evaluation of the terms in the influence action}
\label{appsec:noise}
Here we first consider the integrations in Eq.~(\ref{type1}) which is a typical one out of the six terms of the first type in the evaluation of the 4th order influence action. 
\begin{eqnarray}
    && \int\,d^{4}x_{1}\cdots d^{4}x_{8}\ \mathcal{F}(x_{1},x_{2},x_{9})\mathcal{F}(x_{3},x_{4},x_{10})\mathcal{F}(x_{5},x_{6},x_{11})\mathcal{F}(x_{7},x_{8},x_{12})\nonumber \\
&&\hskip 38pt G(x_{1},x_{3})G(x_{2},x_{8})G(x_{4},x_{6})G(x_{5},x_{7}),
\end{eqnarray}
where $\mathcal{F} (x,x',x'')$ and $G(x,x')$ are defined in Eqs.~(\ref{Fint}) and (\ref{gfct}), respectively. The integrations over $x_{1}$ to $x_{8}$ can be evaluated readily to give
\begin{eqnarray}
&&\int\frac{d^{3}k_{1}}{(2\pi)^{3}}\frac{d^{3}k_{2}}{(2\pi)^{3}}|\vec{k}_{1}|(\hat{\varepsilon}_{R}(\hat{k}_{1})\cdot\hat{\varepsilon}_{R}(\hat{k}_{2}))\int\frac{d^{3}k_{3}}{(2\pi)^{3}}\frac{d^{3}k_{4}}{(2\pi)^{3}}|\vec{k}_{3}|(\hat{\varepsilon}_{R}(\hat{k}_{3})\cdot\hat{\varepsilon}_{R}(\hat{k}_{4}))\nonumber\\
&&\int\frac{d^{3}k_{5}}{(2\pi)^{3}}\frac{d^{3}k_{6}}{(2\pi)^{3}}|\vec{k}_{5}|(\hat{\varepsilon}_{R}(\hat{k}_{5})\cdot\hat{\varepsilon}_{R}(\hat{k}_{6}))\int\frac{d^{3}k_{7}}{(2\pi)^{3}}\frac{d^{3}k_{8}}{(2\pi)^{3}}|\vec{k}_{7}|(\hat{\varepsilon}_{R}(\hat{k}_{7})\cdot\hat{\varepsilon}_{R}(\hat{k}_{8}))\nonumber\\
&&\int\frac{d^{3}k_{9}}{(2\pi)^{3}}g_{k_{9}}(\tau_{9},\tau_{10})\int\frac{d^{3}k_{10}}{(2\pi)^{3}}\partial_{\tau_{9}}\partial_{\tau_{12}}g_{k_{10}}(\tau_{9},\tau_{12})\nonumber\\
&&
\int\frac{d^{3}k_{11}}{(2\pi)^{3}}\partial_{\tau_{10}}\partial_{\tau_{11}}g_{k_{11}}(\tau_{10},\tau_{11})\int\frac{d^{3}k_{12}}{(2\pi)^{3}}g_{k_{12}}(\tau_{11},\tau_{12})\nonumber\\
&&e^{i\vec{k}_{1}\cdot\vec{x}_{9}+i\vec{k}_{2}\cdot\vec{x}_{9}+i\vec{k}_{3}\cdot\vec{x}_{10}+i\vec{k}_{4}\cdot\vec{x}_{10}+i\vec{k}_{5}\cdot\vec{x}_{11}+i\vec{k}_{6}\cdot\vec{x}_{11}+i\vec{k}_{7}\cdot\vec{x}_{12}+i\vec{k}_{8}\cdot\vec{x}_{12}}\nonumber\\
&&(2\pi)^{3}\delta(\vec{k}_{1}-\vec{k}_{9})(2\pi)^{3}\delta(\vec{k}_{2}-\vec{k}_{10})(2\pi)^{3}\delta(\vec{k}_{3}+\vec{k}_{9})(2\pi)^{3}\delta(\vec{k}_{4}-\vec{k}_{11})\nonumber\\
&&(2\pi)^{3}\delta(\vec{k}_{5}-\vec{k}_{12})(2\pi)^{3}\delta(\vec{k}_{6}+\vec{k}_{11})(2\pi)^{3}\delta(\vec{k}_{7}+\vec{k}_{12})(2\pi)^{3}\delta(\vec{k}_{8}+\vec{k}_{10})\nonumber\\
&=&\int\frac{d^{3}k_{9}}{(2\pi)^{3}}\int\frac{d^{3}k_{10}}{(2\pi)^{3}}\int\frac{d^{3}k_{11}}{(2\pi)^{3}}\int\frac{d^{3}k_{12}}{(2\pi)^{3}}\ e^{i\vec{k}_{9}\cdot(\vec{x}_{9}-\vec{x}_{10})+i\vec{k}_{10}\cdot(\vec{x}_{9}-\vec{x}_{12})+i\vec{k}_{11}\cdot(\vec{x}_{10}-\vec{x}_{11})+i\vec{k}_{12}\cdot(\vec{x}_{11}-\vec{x}_{12})}\nonumber\\
&&\hskip 20pt g_{k_{9}}(\tau_{9},\tau_{10})\partial_{\tau_{9}}\partial_{\tau_{12}}g_{k_{10}}(\tau_{9},\tau_{12})\partial_{\tau_{10}}\partial_{\tau_{11}}g_{k_{11}}(\tau_{10},\tau_{11})g_{k_{12}}(\tau_{11},\tau_{12})
\nonumber\\
&&\hskip 20pt |\vec{k}_{9}|(\hat{\varepsilon}_{R}(\hat{k}_{9})\cdot\hat{\varepsilon}_{R}(\hat{k}_{10}))|\vec{k}_{9}|(\hat{\varepsilon}_{R}(-\hat{k}_{9})\cdot\hat{\varepsilon}_{R}(\hat{k}_{11}))\nonumber\\
&&\hskip 20pt |\vec{k}_{12}|(\hat{\varepsilon}_{R}(\hat{k}_{12})\cdot\hat{\varepsilon}_{R}(-\hat{k}_{11}))|\vec{k}_{12}|(\hat{\varepsilon}_{R}(-\hat{k}_{12})\cdot\hat{\varepsilon}_{R}(-\hat{k}_{10})).
\end{eqnarray}
By examining the exponentials, we notice that the momentum dependences are $\vec{k}_{9}+\vec{k}_{10}$, $\vec{k}_{11}+\vec{k}_{10}$, and $\vec{k}_{12}+\vec{k}_{10}$. It is therefore appropriate to make the change of variables: $\vec{K}=\vec{k}_{9}+\vec{k}_{10}$, $\vec{K}'=\vec{k}_{11}+\vec{k}_{10}$, and $\vec{K}''=\vec{k}_{12}+\vec{k}_{10}$. Then, we have
\begin{eqnarray}\label{b2}
&&\int\frac{d^{3}K}{(2\pi)^{3}}\int\frac{d^{3}K'}{(2\pi)^{3}}\int\frac{d^{3}K''}{(2\pi)^{3}}\int\frac{d^{3}k_{10}}{(2\pi)^{3}}\ e^{i\vec{K}\cdot(\vec{x}_{9}-\vec{x}_{10})+i\vec{K'}\cdot(\vec{x}_{10}-\vec{x}_{11})+i\vec{K''}\cdot(\vec{x}_{11}-\vec{x}_{12})}\nonumber\\
&&\hskip 20pt g_{|\vec{K}-\vec{k}_{10}|}(\tau_{9},\tau_{10})\partial_{\tau_{9}}\partial_{\tau_{12}}g_{k_{10}}(\tau_{9},\tau_{12})\partial_{\tau_{10}}\partial_{\tau_{11}}g_{|\vec{K}'-\vec{k}_{10}|}(\tau_{10},\tau_{11})g_{|\vec{K}''-\vec{k}_{10}|}(\tau_{11},\tau_{12})
\nonumber\\
&&\hskip 20pt |\vec{K}-\vec{k}_{10}|(\hat{\varepsilon}_{R}(\hat{K}-\hat{k}_{10})\cdot\hat{\varepsilon}_{R}(\hat{k}_{10}))|\vec{K}-\hat{k}_{10}|(\hat{\varepsilon}_{R}(-\hat{K}+\hat{k}_{10})\cdot\hat{\varepsilon}_{R}(\hat{K}'-\hat{k}_{10}))\nonumber\\
&&\hskip 20pt |\vec{K}''-\vec{k}_{10}|(\hat{\varepsilon}_{R}(\hat{K}''-\hat{k}_{10})\cdot\hat{\varepsilon}_{R}(-\hat{K}'+\hat{k}_{10}))|\vec{K}''-\vec{k}_{10}|(\hat{\varepsilon}_{R}(-\hat{K}''+\hat{k}_{10})\cdot\hat{\varepsilon}_{R}(-\hat{k}_{10})).\nonumber\\
\end{eqnarray}
Next, we work on the integration over $\vec{k}_{10}$. Due to the exponentials in $g_{k}(\tau,\tau')$ as given in Eq.~(\ref{gk}), the dominant part of the integration comes from small values of $|\vec{k}_{10}|$ in the large $\xi$ approximation. Hence, we expand the integrand in Eq.~(\ref{b2}) in powers of $\vec{k}_{10}$ and keep only the lowest order term.
\begin{eqnarray}
&&\int\frac{d^{3}K}{(2\pi)^{3}}\int\frac{d^{3}K'}{(2\pi)^{3}}\int\frac{d^{3}K''}{(2\pi)^{3}}\,K^{2}K''^{2}\ e^{i\vec{K}\cdot(\vec{x}_{9}-\vec{x}_{10})+i\vec{K'}\cdot(\vec{x}_{10}-\vec{x}_{11})+i\vec{K''}\cdot(\vec{x}_{11}-\vec{x}_{12})}\nonumber\\
&&\hskip 20pt g_{K}(\tau_{9},\tau_{10})\partial_{\tau_{10}}\partial_{\tau_{11}}g_{K'}(\tau_{10},\tau_{11})g_{K''}(\tau_{11},\tau_{12})
(\hat{\varepsilon}_{R}(-\hat{K})\cdot\hat{\varepsilon}_{R}(\hat{K}')) (\hat{\varepsilon}_{R}(\hat{K}'')\cdot\hat{\varepsilon}_{R}(-\hat{K}'))\nonumber\\
&&\int\frac{d^{3}k_{10}}{(2\pi)^{3}}\partial_{\tau_{9}}\partial_{\tau_{12}}g_{k_{10}}(\tau_{9},\tau_{12})\,(\hat{\varepsilon}_{R}(\hat{K})\cdot\hat{\varepsilon}_{R}(\hat{k}_{10}))(\hat{\varepsilon}_{R}(-\hat{K}'')\cdot\hat{\varepsilon}_{R}(-\hat{k}_{10})).
\end{eqnarray}
Again, in the large $\xi$ approximation, the $\tau$ derivatives will act on the exponential part of $g_{k}(\tau,\tau')$. Then, the $\vec{k}_{10}$ integration can be done giving
\begin{eqnarray}\label{k10int}
  &&  \int\frac{d^{3}k_{10}}{(2\pi)^{3}}\partial_{\tau_{9}}\partial_{\tau_{12}}g_{k_{10}}(\tau_{9},\tau_{12})\,(\hat{\varepsilon}_{R}(\hat{K})\cdot\hat{\varepsilon}_{R}(\hat{k}_{10}))(\hat{\varepsilon}_{R}(-\hat{K}'')\cdot\hat{\varepsilon}_{R}(-\hat{k}_{10}))\nonumber\\
  &=&e^{2\pi\xi}\left(-\frac{15i}{128\pi^{3}}\right)\left(\frac{1}{\xi^{3}(\tau_{9}\tau_{12})^{1/4}(\sqrt{-\tau_{9}}+\sqrt{-\tau_{12}})^{7}}\right)(\hat{\varepsilon}_{R}(\hat{K})\cdot\hat{\varepsilon}_{R}(-\hat{K}'')),
\end{eqnarray}
where we have used the result 
\begin{eqnarray}
    \int\,d\Omega_{k_{10}}[(\hat{\varepsilon}_{R}(\hat{K})\cdot\hat{\varepsilon}_{R}(\hat{k}_{10})][\hat{\varepsilon}_{R}(-\hat{K}'')\cdot\hat{\varepsilon}_{R}(-\hat{k}_{10}))]=\frac{4}{3}\pi[(\hat{\varepsilon}_{R}(\hat{K})\cdot\hat{\varepsilon}_{R}(-\hat{K}'')].
\end{eqnarray}

Finally, Eq.~(\ref{type1}) can be evaluated to give the leading term in the large $\xi$ expansion:
\begin{eqnarray}
    && \int\,d^{4}x_{1}\cdots d^{4}x_{8}\ \mathcal{F}(x_{1},x_{2},x_{9})\mathcal{F}(x_{3},x_{4},x_{10})\mathcal{F}(x_{5},x_{6},x_{11})\mathcal{F}(x_{7},x_{8},x_{12}) \nonumber \\
&&\hskip 38pt G(x_{1},x_{3})G(x_{2},x_{8})G(x_{4},x_{6})G(x_{5},x_{7})\nonumber\\
&=&\int\frac{d^{3}K}{(2\pi)^{3}}\frac{d^{3}K'}{(2\pi)^{3}}\frac{d^{3}K''}{(2\pi)^{3}}\,e^{i\vec{K}\cdot(\vec{x}_{9}-\vec{x}_{10})}e^{i\vec{K}'\cdot(\vec{x}_{10}-\vec{x}_{11})}e^{i\vec{K}''\cdot(\vec{x}_{11}-\vec{x}_{12})}\nonumber\\
&&\ \ K^{3/2}K'^{1/2}K''^{3/2}\,(\hat{\varepsilon}_{R}(\hat{K})\cdot\hat{\varepsilon}_{R}(-\hat{K}''))(\hat{\varepsilon}_{R}(\hat{K}'')\cdot\hat{\varepsilon}_{R}(-\hat{K}'))(\hat{\varepsilon}_{R}(\hat{K}')\cdot\hat{\varepsilon}_{R}(-\hat{K}))\nonumber\\
&&\ \ \left(\frac{e^{8\pi\xi}}{\xi^{7/2}}\right)\left(\frac{15\sqrt{2}}{2048\pi^{2}}\right)\frac{1}{(\sqrt{-\tau_{9}}+\sqrt{-\tau_{12}})^{7}}\nonumber\\
&&\ \ e^{-2\sqrt{2\xi K}(\sqrt{-\tau_{9}}+\sqrt{-\tau_{10}})}e^{-2\sqrt{2\xi K'}(\sqrt{-\tau_{10}}+\sqrt{-\tau_{11}})}e^{-2\sqrt{2\xi K''}(\sqrt{-\tau_{11}}+\sqrt{-\tau_{12}})}\,.
\end{eqnarray}

The evaluation of the integrations for the term of the second type in Eq.~(\ref{type2}) can be done along the same lines as above for the first type of terms. After integrations over $x_{1}$ to $x_{8}$, we have
\begin{eqnarray}
    && \int\,d^{4}x_{1}\cdots d^{4}x_{8}\ \mathcal{F}(x_{1},x_{2},x_{9})\mathcal{F}(x_{3},x_{4},x_{10})\mathcal{F}(x_{5},x_{6},x_{11})\mathcal{F}(x_{7},x_{8},x_{12}) \nonumber \\
&&\hskip 38pt G(x_{1},x_{3})G(x_{2},x_{8})G(x_{4},x_{5})G(x_{6},x_{7}) \nonumber\\
&=&\int\frac{d^{3}k_{9}}{(2\pi)^{3}}\int\frac{d^{3}k_{10}}{(2\pi)^{3}}\int\frac{d^{3}k_{11}}{(2\pi)^{3}}\int\frac{d^{3}k_{12}}{(2\pi)^{3}}\ e^{i\vec{k}_{9}\cdot(\vec{x}_{9}-\vec{x}_{10})+i\vec{k}_{10}\cdot(\vec{x}_{9}-\vec{x}_{12})+i\vec{k}_{11}\cdot(\vec{x}_{10}-\vec{x}_{11})+i\vec{k}_{12}\cdot(\vec{x}_{11}-\vec{x}_{12})}\nonumber\\
&&\hskip 20pt g_{k_{9}}(\tau_{9},\tau_{10})\partial_{\tau_{9}}\partial_{\tau_{12}}g_{k_{10}}(\tau_{9},\tau_{12})\partial_{\tau_{10}}g_{k_{11}}(\tau_{10},\tau_{11})\partial_{\tau_{11}}g_{k_{12}}(\tau_{11},\tau_{12})
\nonumber\\
&&\hskip 20pt |\vec{k}_{9}|(\hat{\varepsilon}_{R}(\hat{k}_{9})\cdot\hat{\varepsilon}_{R}(\hat{k}_{10}))|\vec{k}_{9}|(\hat{\varepsilon}_{R}(-\hat{k}_{9})\cdot\hat{\varepsilon}_{R}(\hat{k}_{11}))\nonumber\\
&&\hskip 20pt |\vec{k}_{11}|(\hat{\varepsilon}_{R}(\hat{k}_{12})\cdot\hat{\varepsilon}_{R}(-\hat{k}_{11}))|\vec{k}_{12}|(\hat{\varepsilon}_{R}(-\hat{k}_{12})\cdot\hat{\varepsilon}_{R}(-\hat{k}_{10})).
\end{eqnarray}
After changing variables to $\vec{K}=\vec{k}_{9}+\vec{k}_{10}$, $\vec{K}'=\vec{k}_{11}+\vec{k}_{10}$, and $\vec{K}''=\vec{k}_{12}+\vec{k}_{10}$, we expand the result in powers of $\vec{k}_{10}$ in the large $\xi$ approximation and retain only the leading term to have
\begin{eqnarray}
&&\int\frac{d^{3}K}{(2\pi)^{3}}\int\frac{d^{3}K'}{(2\pi)^{3}}\int\frac{d^{3}K''}{(2\pi)^{3}}\,K^{2}K'K''\ e^{i\vec{K}\cdot(\vec{x}_{9}-\vec{x}_{10})+i\vec{K'}\cdot(\vec{x}_{10}-\vec{x}_{11})+i\vec{K''}\cdot(\vec{x}_{11}-\vec{x}_{12})}\nonumber\\
&&\hskip 20pt g_{K}(\tau_{9},\tau_{10})\partial_{\tau_{10}}g_{K'}(\tau_{10},\tau_{11})\partial_{\tau_{11}}g_{K''}(\tau_{11},\tau_{12})
(\hat{\varepsilon}_{R}(-\hat{K})\cdot\hat{\varepsilon}_{R}(\hat{K}')) (\hat{\varepsilon}_{R}(\hat{K}'')\cdot\hat{\varepsilon}_{R}(-\hat{K}'))\nonumber\\
&&\int\frac{d^{3}k_{10}}{(2\pi)^{3}}\partial_{\tau_{9}}\partial_{\tau_{12}}g_{k_{10}}(\tau_{9},\tau_{12})\,(\hat{\varepsilon}_{R}(\hat{K})\cdot\hat{\varepsilon}_{R}(\hat{k}_{10}))(\hat{\varepsilon}_{R}(-\hat{K}'')\cdot\hat{\varepsilon}_{R}(-\hat{k}_{10})).
\end{eqnarray}
The integration over $k_{10}$ is the same as in Eq.~(\ref{k10int}). In addition, in the large $\xi$ approximation, the leading term will have the $\tau$ derivative acting only on the exponential of $g_{k}(\tau,\tau')$. Hence, Eq.~(\ref{type2}) can be simplified to give
\begin{eqnarray}
    && \int\,d^{4}x_{1}\cdots d^{4}x_{8}\ \mathcal{F}(x_{1},x_{2},x_{9})\mathcal{F}(x_{3},x_{4},x_{10})\mathcal{F}(x_{5},x_{6},x_{11})\mathcal{F}(x_{7},x_{8},x_{12}) \nonumber \\
&&\hskip 38pt G(x_{1},x_{3})G(x_{2},x_{8})G(x_{4},x_{5})G(x_{6},x_{7})\nonumber\\
&=&\int\frac{d^{3}K}{(2\pi)^{3}}\frac{d^{3}K'}{(2\pi)^{3}}\frac{d^{3}K''}{(2\pi)^{3}}\,e^{i\vec{K}\cdot(\vec{x}_{9}-\vec{x}_{10})}e^{i\vec{K}'\cdot(\vec{x}_{10}-\vec{x}_{11})}e^{i\vec{K}''\cdot(\vec{x}_{11}-\vec{x}_{12})}\nonumber\\
&&\ \ K^{3/2}K'K''\,(\hat{\varepsilon}_{R}(\hat{K})\cdot\hat{\varepsilon}_{R}(-\hat{K}''))(\hat{\varepsilon}_{R}(\hat{K}'')\cdot\hat{\varepsilon}_{R}(-\hat{K}'))(\hat{\varepsilon}_{R}(\hat{K}')\cdot\hat{\varepsilon}_{R}(-\hat{K}))\nonumber\\
&&\ \ \left(\frac{e^{8\pi\xi}}{\xi^{7/2}}\right)\left(\frac{15\sqrt{2}}{2048\pi^{2}}\right)\frac{1}{(\sqrt{-\tau_{9}}+\sqrt{-\tau_{12}})^{7}}\nonumber\\
&&\ \ e^{-2\sqrt{2\xi K}(\sqrt{-\tau_{9}}+\sqrt{-\tau_{10}})}e^{-2\sqrt{2\xi K'}(\sqrt{-\tau_{10}}+\sqrt{-\tau_{11}})}e^{-2\sqrt{2\xi K''}(\sqrt{-\tau_{11}}+\sqrt{-\tau_{12}})}\,.
\end{eqnarray}

Lastly, we also indicate how to evaluate Eq.~(\ref{type3}) corresponding to one of terms of the third type. Integating over $x_{1}$ to $x_{8}$, 
\begin{eqnarray}
    && \int\,d^{4}x_{1}\cdots d^{4}x_{8}\ \mathcal{F}(x_{1},x_{2},x_{9})\mathcal{F}(x_{3},x_{4},x_{10})\mathcal{F}(x_{5},x_{6},x_{11})\mathcal{F}(x_{7},x_{8},x_{12}) \nonumber \\
&&\hskip 38pt G(x_{1},x_{4})G(x_{2},x_{7})G(x_{3},x_{6})G(x_{5},x_{8}) \nonumber\\
&=&\int\frac{d^{3}k_{9}}{(2\pi)^{3}}\int\frac{d^{3}k_{10}}{(2\pi)^{3}}\int\frac{d^{3}k_{11}}{(2\pi)^{3}}\int\frac{d^{3}k_{12}}{(2\pi)^{3}}\ e^{i\vec{k}_{9}\cdot(\vec{x}_{9}-\vec{x}_{10})+i\vec{k}_{10}\cdot(\vec{x}_{9}-\vec{x}_{12})+i\vec{k}_{11}\cdot(\vec{x}_{10}-\vec{x}_{11})+i\vec{k}_{12}\cdot(\vec{x}_{11}-\vec{x}_{12})}\nonumber\\
&&\hskip 20pt \partial_{\tau_{10}}g_{k_{9}}(\tau_{9},\tau_{10})\partial_{\tau_{9}}g_{k_{10}}(\tau_{9},\tau_{12})\partial_{\tau_{11}}g_{k_{11}}(\tau_{10},\tau_{11})\partial_{\tau_{12}}g_{k_{12}}(\tau_{11},\tau_{12})
\nonumber\\
&&\hskip 20pt |\vec{k}_{9}|(\hat{\varepsilon}_{R}(\hat{k}_{9})\cdot\hat{\varepsilon}_{R}(\hat{k}_{10}))|\vec{k}_{11}|(\hat{\varepsilon}_{R}(\hat{k}_{11})\cdot\hat{\varepsilon}_{R}(-\hat{k}_{9}))\nonumber\\
&&\hskip 20pt |\vec{k}_{12}|(\hat{\varepsilon}_{R}(\hat{k}_{12})\cdot\hat{\varepsilon}_{R}(-\hat{k}_{11}))|\vec{k}_{10}|(\hat{\varepsilon}_{R}(-\hat{k}_{10})\cdot\hat{\varepsilon}_{R}(-\hat{k}_{12})).
\end{eqnarray}
Working through similar steps as above, we change variables to $\vec{K}=\vec{k}_{9}+\vec{k}_{10}$, $\vec{K}'=\vec{k}_{11}+\vec{k}_{10}$, and $\vec{K}''=\vec{k}_{12}+\vec{k}_{10}$, and expand the result in powers of $\vec{k}_{10}$. After doing the integration over $k_{10}$, we obtain the leading $\xi$ contribution for this term of the third type in Eq.~(\ref{type3}),
\begin{eqnarray}
    && \int\,d^{4}x_{1}\cdots d^{4}x_{8}\ \mathcal{F}(x_{1},x_{2},x_{9})\mathcal{F}(x_{3},x_{4},x_{10})\mathcal{F}(x_{5},x_{6},x_{11})\mathcal{F}(x_{7},x_{8},x_{12}) \nonumber \\
&&\hskip 38pt G(x_{1},x_{4})G(x_{2},x_{7})G(x_{3},x_{6})G(x_{5},x_{8})\nonumber\\
&=&\int\frac{d^{3}K}{(2\pi)^{3}}\frac{d^{3}K'}{(2\pi)^{3}}\frac{d^{3}K''}{(2\pi)^{3}}\,e^{i\vec{K}\cdot(\vec{x}_{9}-\vec{x}_{10})}e^{i\vec{K}'\cdot(\vec{x}_{10}-\vec{x}_{11})}e^{i\vec{K}''\cdot(\vec{x}_{11}-\vec{x}_{12})}\nonumber\\
&&\ \ KK'K''\,(\hat{\varepsilon}_{R}(\hat{K})\cdot\hat{\varepsilon}_{R}(-\hat{K}''))(\hat{\varepsilon}_{R}(\hat{K}'')\cdot\hat{\varepsilon}_{R}(-\hat{K}'))(\hat{\varepsilon}_{R}(\hat{K}')\cdot\hat{\varepsilon}_{R}(-\hat{K}))\nonumber\\
&&\ \ \left(\frac{e^{8\pi\xi}}{\xi^{4}}\right)\left(\frac{105}{4096\pi^{2}}\right)\frac{1}{(\sqrt{-\tau_{9}}+\sqrt{-\tau_{12}})^{8}}\nonumber\\
&&\ \ e^{-2\sqrt{2\xi K}(\sqrt{-\tau_{9}}+\sqrt{-\tau_{10}})}e^{-2\sqrt{2\xi K'}(\sqrt{-\tau_{10}}+\sqrt{-\tau_{11}})}e^{-2\sqrt{2\xi K''}(\sqrt{-\tau_{11}}+\sqrt{-\tau_{12}})}\,.
\end{eqnarray}

\section{Derivation of the four-point correlation function in Eq.~(\ref{4ptcf1})}
\label{appsec:der4pt}

In this appendix, we detail the derivation of Eq.~(\ref{4ptcf1}). 
\begin{eqnarray}\label{con4pt}
&&\langle \phi(\tau,\vec{x})\phi(\tau,\vec{x}')\phi(\tau,\vec{x}'')\phi(\tau,\vec{x}''')\rangle _{conn}\nonumber \\
&=& \int d^4x_1 \sqrt{-g(x_1)} \int d^4x_2 \sqrt{-g(x_2)} \int d^4x_3 \sqrt{-g(x_3)}  \int d^4x_4 \sqrt{-g(x_4)} \times \nonumber \\
&&\ \ \ \ \  G_{\rm ret}^\phi(x,x_1) G_{\rm ret}^\phi(x',x_2) G_{\rm ret}^\phi(x'',x_3)G_{\rm ret}^\phi(x''',x_4) N_4(x_1,x_2,x_3,x_4) {\Big |}_{\tau>\tau'\rightarrow\tau>\tau''\rightarrow\tau>\tau'''\rightarrow\tau}\,.\nonumber\\
\end{eqnarray}
The retarded Green's function $G_{ret}^{\phi}$ is given in Eq.~(\ref{retG}), and it can be expressed more explicitly as
\begin{eqnarray}
    G_{ret}^{\phi}(x,x')&=&i\theta(\tau-\tau')\left(\frac{1}{a(\tau)a(\tau')}\right)\int\,\frac{d^{3}k}{(2\pi)^{3}}\,g_{ret}(k,\tau,\tau')e^{i\vec{k}\cdot(\vec{x}-\vec{x}')},\\
    g_{ret}(k,\tau,\tau')&=&\frac{1}{2k}\left[\left(1-\frac{i}{k\tau}\right)\left(1+\frac{i}{k\tau'}\right)e^{-ik(\tau-\tau')}-\left(1+\frac{i}{k\tau}\right)\left(1-\frac{i}{k\tau'}\right)e^{ik(\tau-\tau')}\right].\nonumber\\
\end{eqnarray}
$N_{4}(x_1,x_2,x_3,x_4)$ is given in Eq.~(\ref{Noise4}).

The integrations over $\vec{x}_{1}$ to $\vec{x}_{4}$ as well as the $k$-integrations over $G_{ret}$ in Eq.~(\ref{con4pt}) can be done readily to give
\begin{eqnarray}
    &&\frac{1}{a^4(\tau)a^4(\tau')a^4(\tau'')a^4(\tau''')}
\left(\frac{\alpha^{4}}{f^{4}}e^{8\pi\xi}\right)\nonumber\\
&&\int\frac{d^{3}k}{(2\pi)^{3}}\int\frac{d^{3}k'}{(2\pi)^{3}}\int\frac{d^{3}k''}{(2\pi)^{3}}\,k k' k''(\hat{\varepsilon}_{R}(\hat{k})\cdot\hat{\varepsilon}_{R}(-\hat{k}''))(\hat{\varepsilon}_{R}(\hat{k}'')\cdot\hat{\varepsilon}_{R}(-\hat{k}'))(\hat{\varepsilon}_{R}(\hat{k}')\cdot\hat{\varepsilon}_{R}(-\hat{k}))\nonumber\\
&&\ \ \ \ e^{i\vec{k}\cdot\vec{x}}\,e^{-i(\vec{k}-\vec{k}')\cdot\vec{x}'}e^{-i(\vec{k}'-\vec{k}'')\cdot\vec{x}''}e^{-i\vec{k}''\cdot\vec{x}'''}\nonumber\\
&&\ \ \ \ \int_{-\infty}^{\tau}d\tau_{1}\int_{-\infty}^{\tau'}d\tau_{2}\int_{-\infty}^{\tau''}d\tau_{3}\int_{-\infty}^{\tau'''}d\tau_{4}\
\frac{1}{a^4(\tau_1)a^4(\tau_2)a^4(\tau_3)a^4(\tau_4)}\nonumber\\
&&\ \ \ \ \left[\left( \frac{15\sqrt{2}}{8192\pi^2\xi^{7/2}}\right)
\frac{(6k^{1/2}+k^{1/2}k'^{-1/2}k''^{1/2})}{(\sqrt{-\tau_1}+\sqrt{-\tau_3})^{7}}+\left(\frac{105}{16384\pi^{2}\xi^{4}}\right)\frac{1}{(\sqrt{-\tau_1}+\sqrt{-\tau_3})^{8}}\right]\nonumber\\
&&\ \ \ \ e^{-2\sqrt{2\xi k}(\sqrt{-\tau_{1}}+\sqrt{-\tau_{2}})}e^{-2\sqrt{2\xi k'}(\sqrt{-\tau_{2}}+\sqrt{-\tau_{3}})}e^{-2\sqrt{2\xi k''}(\sqrt{-\tau_{3}}+\sqrt{-\tau_{4}})}\nonumber\\
&&\ \ \ \ g_{ret}(k,\tau,\tau_{1})g_{ret}(|\vec{k}-\vec{k}'|,\tau',\tau_{2})g_{ret}(|\vec{k}'-\vec{k}''|,\tau'',\tau_{3})g_{ret}(k,\tau''',\tau_{4})\nonumber\\
&&\hskip 100pt \ +\;23\ {\rm permutations\ of\ }(x_{1},x_{2},x_{3},x_{4})\,.
\end{eqnarray}

To evaluate the integrations over $\tau$'s, we consider the large $\xi$ approximation of the following integral,
\begin{eqnarray}
    I(a,b,c,A,B)\equiv\int_{-\infty}^{\tau}d\tau_{1}(-\tau_{1})^{a}\,g_{ret}(k,\tau,\tau_{1})\frac{1}{(A\sqrt{-\tau_{1}}+B)^{b}}\,e^{-c\sqrt{\xi(-\tau_{1})}}\,.
\end{eqnarray}
For large $\xi$, the main contribution of the integral comes from small values of $\tau_{1}$. In particular, we expand $\sqrt{-\tau_{1}}$  in $g_{ret}(k,\tau,\tau_{1})$ around $\sqrt{-\tau}$, so we have
\begin{eqnarray}
    g_{ret}(k,\tau,\tau_{1})=-2i\sqrt{-\tau}\,(\sqrt{-\tau_{1}}-\sqrt{-\tau})+\cdots
\end{eqnarray}
With this approximation, $I(a,b,c,A,B)$ can be evaluated as
\begin{eqnarray}
    I(a,b,c,A,B)=\frac{(-4i)(-\tau)^{a+1}}{(A\sqrt{-\tau}+B)^{b}c^{2}\xi}\,e^{-c\sqrt{\xi(-\tau)}}+\cdots
\end{eqnarray}
Using this approximation for $I(a,b,c,A,B)$, the result of the $\tau$ integrations in the large $\xi$ limit can be expressed as
\begin{eqnarray}
    &&\ \ \ \ \int_{-\infty}^{\tau}d\tau_{1}\int_{-\infty}^{\tau'}d\tau_{2}\int_{-\infty}^{\tau''}d\tau_{3}\int_{-\infty}^{\tau'''}d\tau_{4}\
\frac{1}{a^4(\tau_1)a^4(\tau_2)a^4(\tau_3)a^4(\tau_4)}\nonumber\\
&&\ \ \ \ \left[\left( \frac{15\sqrt{2}}{8192\pi^2\xi^{7/2}}\right)
\frac{(6k^{1/2}+k^{1/2}k'^{-1/2}k''^{1/2})}{(\sqrt{-\tau_1}+\sqrt{-\tau_3})^{7}}+\left(\frac{105}{16384\pi^{2}\xi^{4}}\right)\frac{1}{(\sqrt{-\tau_1}+\sqrt{-\tau_3})^{8}}\right]\nonumber\\
&&\ \ \ \ e^{-2\sqrt{2\xi k}(\sqrt{-\tau_{1}}+\sqrt{-\tau_{2}})}e^{-2\sqrt{2\xi k'}(\sqrt{-\tau_{2}}+\sqrt{-\tau_{3}})}e^{-2\sqrt{2\xi k''}(\sqrt{-\tau_{3}}+\sqrt{-\tau_{4}})}\nonumber\\
&&\ \ \ \ g_{ret}(k,\tau,\tau_{1})g_{ret}(|\vec{k}-\vec{k}'|,\tau',\tau_{2})g_{ret}(|\vec{k}'-\vec{k}''|,\tau'',\tau_{3})g_{ret}(k,\tau''',\tau_{4})\nonumber\\
&=&\left(\frac{H^{4}}{16\xi^{4}}\right)\left[\frac{1}{kk''(\sqrt{k}+\sqrt{k'})^{2}(\sqrt{k'}+\sqrt{k''})^{2}}\right](-\tau)^{2}(-\tau')^{2}(-\tau'')^{2}(-\tau''')^{2}\nonumber\\
&&\ \ e^{-2\sqrt{2\xi k}(\sqrt{-\tau}+\sqrt{-\tau'})}e^{-2\sqrt{2\xi k'}(\sqrt{-\tau'}+\sqrt{-\tau''})}e^{-2\sqrt{2\xi k''}(\sqrt{-\tau''}+\sqrt{-\tau'''})}\nonumber\\
&&\ \ \left[\left( \frac{15\sqrt{2}}{8192\pi^2\xi^{7/2}}\right)\frac{(6k^{1/2}+k^{1/2}k'^{-1/2}k''^{1/2})}{(\sqrt{-\tau}+\sqrt{-\tau'''})^{7}}+\left(\frac{105}{16384\pi^{2}\xi^{4}}\right)\frac{1}{(\sqrt{-\tau_1}+\sqrt{-\tau_3})^{8}}\right]+\cdots\nonumber\\
\end{eqnarray}
Finally, taking the limits $\tau'$, $\tau''$, $\tau'''\rightarrow\tau$, we simplify the connected four-point correlation function in Eq.~(\ref{con4pt}) to
\begin{eqnarray}
&&\left\langle\phi(\tau,\vec{x})\phi(\tau,\vec{x}')\phi(\tau,\vec{x}'')\phi(\tau,\vec{x}''')\right\rangle_{conn}\nonumber\\
    &=&\left(\frac{15}{2^{26}\pi^{2}}\right)\left(\frac{\alpha^{4}}{f^{4}}e^{8\pi\xi}\right)\left(\frac{H(-\tau)}{\xi}\right)^{8}\nonumber\\
    &&\int\frac{d^{3}k}{(2\pi)^{3}}\int\frac{d^{3}k'}{(2\pi)^{3}}\int\frac{d^{3}k''}{(2\pi)^{3}}
    (\hat{\epsilon}_{R}(\hat{k})\cdot\hat{\epsilon}_{R}(-\hat{k}''))
    (\hat{\epsilon}_{R}(\hat{k}'')\cdot\hat{\epsilon}_{R}(-\hat{k}'))(\hat{\epsilon}_{R}(\hat{k}')\cdot\hat{\epsilon}_{R}(-\hat{k}))\nonumber\\
    &&\ \ \left[\frac{e^{-4\sqrt{2\xi(-\tau)}(\sqrt{k}+\sqrt{k'}+\sqrt{k''})}}{(\sqrt{k}+\sqrt{k'})^{2}(\sqrt{k'}+\sqrt{k''})^{2}}\right]
    \Big[7k'+4\sqrt{2\xi(-\tau)}(k^{1/2}k'^{1/2})(6k'^{1/2}+k''^{1/2})\Big]\nonumber\\
    &&\ \ \left\{e^{i\vec{k}\cdot\vec{x}}e^{-i(\vec{k}-\vec{k}')\cdot\vec{x}'}e^{-i(\vec{k}'-\vec{k}'')\cdot\vec{x}''}e^{-i\vec{k}''\cdot\vec{x}''}+23\ {\rm permutations\ of }\ \vec{x},\vec{x}',\vec{x}'',\vec{x}'''\right\}.
\end{eqnarray}

\section{Wigner-3j symbols}
\label{sec:3j}

We adopt the spherical harmonics given by
\begin{equation}
Y_{\ell m}(\hat{n})=\sqrt{\frac{(2\ell+1)}{(4\pi)}\frac{(\ell-m)!}{(\ell+m)!}}P^m_\ell(\cos \theta) e^{i m \phi} \,,
\end{equation}
which satisfy the orthogonal relation,
\begin{equation}\label{YYint}
    \int d{\hat{n}}\; Y^*_{\ell m}(\hat{n}) Y_{\ell' m'}(\hat{n})
    = \delta_{\ell \ell'} \delta_{m m'} \,,
\end{equation}
and the completeness relation,
\begin{equation}
    \sum_{\ell m} Y^*_{\ell m}(\hat{n}) Y_{\ell m}(\hat{n}')
    = \delta(\hat{n}-\hat{n}')
    = \delta(\phi-\phi')\delta(\cos\theta-\cos\theta')\,.
\end{equation}
Its complex conjugate is
\begin{equation}
Y^*_{\ell m}(\hat{n}) =  (-1)^{m} Y_{\ell -m}(\hat{n}) \,,
\end{equation}
and its parity is given by
\begin{equation}
Y_{\ell m}(-\hat{n}) \equiv Y_{\ell m}(\pi-\theta,\phi+\pi)=(-1)^{\ell} Y_{\ell m}(\hat{n}) \,.
\end{equation}

We can calculate the integral of a product of three spherical harmonics using the formula:
\begin{eqnarray}\label{YYYint}
    \int d\hat{e} \;
   Y_{\ell_1 m_1}(\hat{e})\; Y_{\ell_2 m_2}(\hat{e})\; Y_{\ell_3 m_3}(\hat{e})&=&
    \sqrt{\frac{(2\ell_1+1)(2\ell_2+1)(2\ell_3+1)}{4\pi}} \times \nonumber \\
   && \six{\ell_1}{\ell_2}{\ell_3}{0}{0}{0}
    \six{\ell_1}{\ell_2}{\ell_3}{m_1}{m_2}{m_3} \,,
\end{eqnarray}
which involves two Wigner-3j symbols representing the coupling coefficients between different spherical harmonics. The Wigner-3j symbol is zero unless it satisfies: $\ell_1$, $\ell_2$, and $\ell_3$ have to meet the triangular condition, i.e.~$\ell_1 + \ell_2 \ge \ell_3 \ge |\ell_1- \ell_2|$, while $m_1+m_2+m_3=0$; when $m_1=m_2=m_3=0$, $\ell_1+\ell_2+\ell_3$ is even.  The Wigner-3j symbols have the reflection property and the summation relation:
\begin{equation}
 \six{\ell_1}{\ell_2}{\ell_3}{m_1}{m_2}{m_3}
  = (-1)^{\ell_1 + \ell_2 +  \ell_3}
\six{\ell_1}{\ell_2}{\ell_3}{-m_1}{-m_2}{-m_3}
  = (-1)^{\ell_1 + \ell_2 +  \ell_3}
 \six{\ell_1}{\ell_3}{\ell_2}{m_1}{m_3}{m_2}\,,
\end{equation}
\begin{equation}
(2\ell +1)\sum_{m_1 m_2}
\six{\ell}{\ell_1}{\ell_2}{m}{m_1}{m_2}
\six{\ell'}{\ell_1}{\ell_2}{m'}{m_1}{m_2}
=  \delta_{\ell \ell'} \delta_{m m'}\,.
\end{equation}

\section{Functions ${\cal R}_{l_{1}\cdots l_{6}}$ and ${\cal E}_{l_{1}m_{1}\cdots l_{6}m_{6}}$}
\label{REvalues}
In this appendix, we work out in details the functions ${\cal R}_{l_1\cdots l_6}$ and ${\cal E}_{l_{1}m_{1}\cdots l_{6}m_{6}}$ defined in Eqs.~(\ref{calR}) and (\ref{calE}), respectively. 

First, for the function ${\cal R}_{l_1\cdots l_6}$, we make changes of variables:
\begin{eqnarray}
    \eta=4\sqrt{k}\ \ \ ,\ \ \ \eta'=4\sqrt{k'}\ \ \ ,\ \ \ \eta''=4\sqrt{k''}.
\end{eqnarray}
This function can be rewritten as
\begin{eqnarray}\label{calReta}
    {\cal R}_{l_1\cdots l_6}&=&2^{-29-4(l_1+\cdots l_6)}\prod_{i=1}^{6}\frac{1}{(2l_{i}+1)!!}\nonumber\\
    &&\int_{0}^{\infty}d\eta\,\eta^{\,5+2(l_1+l_2)}e^{-\eta}\int_{0}^{\infty}d\eta'\,\eta'^{\,5+2(l_3+l_4)}e^{-\eta'}\int_{0}^{\infty}d\eta''\,\eta''^{\,5+2(l_5+l_6)}e^{-\eta''}\nonumber\\
    &&\ \ (\eta+\eta')^{-2}(\eta'+\eta'')^{-2}[7\eta'^{2}+\eta\eta'(6\eta'+\eta'')]\,.
\end{eqnarray}
Using the identity,
\begin{eqnarray}(\eta+\eta')^{-2}=\int_{0}^{\infty}d\tau\,\tau\,e^{-\tau(\eta+\eta')}\,.
\end{eqnarray}
The integrations over $\eta$, $\eta'$ and $\eta''$ can be worked out to give
\begin{eqnarray}\label{calRfinal}
    {\cal R}_{l_1\cdots l_6}&=&2^{-29-4(l_1+\cdots l_6)}\prod_{i=1}^{6}\frac{1}{(2l_{i}+1)!!}
    \Gamma[6+2(l_1+l_2)]\,\Gamma[7+2(l_3+l_4)]\,\Gamma[6+2(l_5+l_6)]\nonumber\\
    &&\int_{0}^{\infty}d\tau\int_{0}^{\infty}d\tau'\,\tau\tau'\,(1+\tau)^{-7-2(l_1+l_2)}(1+\tau')^{-7-2(l_3+l_4)}(1+\tau+\tau')^{-2[4+(l_5+l_6)]}\nonumber\\
    &&\{7(7+2(l_3+l_4))(1+\tau)(1+
    \tau')\nonumber\\
    &&+4(3+(l_1+l_2))[(l_5+l_6)(1+\tau+\tau')+3(1+\tau)+(24+6(l_3+l_4))(1+\tau')-3]\}.\nonumber\\
\end{eqnarray}
In this form, it is easy to evaluate the integrals over $\tau$ and $\tau'$ for a particular set of $l$'s. For example, for ${l_1,\cdots,l_6}={0,\cdots,0}$, we have
\begin{eqnarray}\label{R0}
    {\cal R}_{000000}&=&2^{-29}\Gamma[6]^2\Gamma[7]\int_{0}^{\infty}d\tau\int_{0}^{\infty}d\tau'\,\tau\tau'(1+\tau)^{-7}(1+\tau')^{-7}(1+\tau+\tau')^{-8}\nonumber\\
    &&\hskip 80pt [49(1+\tau)(1+\tau')+36(1+\tau)+288(1+\tau')-36]\nonumber\\
    &=&\frac{301659988953375}{5767168}-\frac{2778595194375\pi^{2}}{524288}\nonumber\\
    &\approx&0.000283213.
\end{eqnarray}
The values of ${\cal R}$ for other sets of $l$'s can be obtained in a similar fashion. For example,
\begin{eqnarray}\label{R1}
    {\cal R}_{010101}&=&\frac{841884090694470546975}{19193135104}-\frac{74563542446518125\pi^2}{16777216}\nonumber\\
    &\approx&0.000135246.
\end{eqnarray}
Since ${\cal R}$ depends only on $\ell_1+\ell_2$, $\ell_3+\ell_4$, and $\ell_5+\ell_6$ divided by $(2\ell_{i}+1)!!$, we have ${\cal R}_{101010}={\cal R}_{010101}$ and so on. Some other numerical values of ${\cal R}_{\ell_{1}\cdots\ell_{6}}$ are tabulated in Table~\ref{Rtable}.

\begin{center}
\begin{longtable}{|c|l|c|}
\caption{Numerical values of ${\cal R}_{\ell_{1}\cdots\ell_{6}}$. \label{Rtable} } \\ 
\hline
\hline 
$\ell_{1}\cdots\ell_{6}$ & Exact value & Approximate \\
& & numerical value \\
\hline
\endfirsthead

\multicolumn{3}{c}%
{\tablename\ \thetable{} -- continued from previous page} \\
\hline 
$\ell_{1}\cdots\ell_{6}$ & Exact value & Approximate \\
& & numerical value \\
\hline 
\endhead

\hline \multicolumn{3}{|r|}{{Continue on next page}} \\ \hline 
\endfoot

\hline \hline
\endlastfoot

000000 & $\frac{301659988953375}{5767168}-\frac{2778595194375\pi^{2}}{524288}$ & 0.000283213 \\
010101 & $\frac{841884090694470546975}{19193135104}-\frac{74563542446518125\pi^2}{16777216}$ & 0.000135246 \\
011124 & $\frac{361382300218100292225709563398835}{3843892650704896} - \frac{
 327299792669198557117284375 \pi^2}{34359738368}$ & 0.001251183 \\
011133 & $\frac{464634385994700375718769438655645}{3843892650704896} - \frac{
 420814019146112430579365625 \pi^2}{34359738368}$ &  0.001608664 \\
 011223 & $\frac{6097776012970397491635193970763549}{32673087530991616} - \frac{
 649727717412353480411135625 \pi^2}{34359738368}$ & 0.001010856 \\
 011313 & $\frac{9523432618005515017501696123847925}{65346175061983232} - \frac{
 507368434300106827150828125 \pi^2}{34359738368}$ & 0.000582114 \\
 012222 & $\frac{18665927931290809434303324402741933}{65346175061983232} - \frac{
 994442131228209381215623125 \pi^2}{34359738368}$ & 0.001140943 \\
 012312 & $\frac{3617514826118661766442624776871883}{16336543765495808} - \frac{
 770903898627945635611183125 \pi^2}{34359738368}$ & 0.001172085 \\
 012402 & $\frac{2601218823956928851189788263286143}{32673087530991616} - \frac{
 277163996411913904403735625 \pi^2}{34359738368}$ & 0.001020381 \\
 013302 & $\frac{3344424202230337094386870624225041}{32673087530991616} - \frac{
 356353709672460734233374375 \pi^2}{34359738368}$ & 0.001311919 \\
 020124 & $\frac{201837335922068306618335326685767}{3843892650704896} - \frac{
 182801753600901934205581875 \pi^2}{34359738368}$ & 0.000720582 \\
 020133 & $\frac{259505146185516394223573991453129}{3843892650704896} - \frac{
 235030826058302486835748125 \pi^2}{34359738368}$ & 0.000926462 \\
 020223 & $\frac{34040150282929023249104677467710601}{326730875309916160} - \frac{
 362703206819287969125616125 \pi^2}{34359738368}$ & 0.000404250 \\
 021222 & $\frac{3773830683890886508708613493507963}{14205690230865920} - \frac{
 924847600032889065589431375 \pi^2}{34359738368}$ & 0.000559701 \\
 021312 & $\frac{13452742018878055974213755571699057}{65346175061983232} - \frac{
 716705512491028240394589375 \pi^2}{34359738368}$ & 0.000441321 \\
 022302 & $\frac{21758631971743370129762422378545561}{163365437654958080} - \frac{
 463683357832281431792322375 \pi^2}{34359738368}$ & 0.000547903 \\
 200124 & $\frac{201837335922068306618335326685767}{3843892650704896} - \frac{
 182801753600901934205581875 \pi^2}{34359738368}$ & 0.000720582 \\
 200133 & $\frac{259505146185516394223573991453129}{3843892650704896} - \frac{
 235030826058302486835748125 \pi^2}{34359738368}$ & 0.000926462 \\
 200223 & $\frac{34040150282929023249104677467710601}{326730875309916160} - \frac{
 362703206819287969125616125 \pi^2}{34359738368}$ & 0.000404250 \\
 203220 & $\frac{21758631971743370129762422378545561}{163365437654958080} - \frac{
 463683357832281431792322375 \pi^2}{34359738368}$ & 0.000547903 \\
 203310 & $\frac{3334021559309950899775718337578673}{32673087530991616} - \frac{
 355245291549961370962779375 \pi^2}{34359738368}$ & 0.001580925 \\
 204210 & $\frac{2593127879463295144270003151450079}{32673087530991616} - \frac{
 276301893427747732971050625 \pi^2}{34359738368}$ & 0.001229608 \\
 213120 & $\frac{13404196351916253732658735382871645}{65346175061983232} - \frac{
 714119203538529726096534375 \pi^2}{34359738368}$ & 0.000513355 \\
 213210 & $\frac{3593241992637760645670466337189677}{16336543765495808} - \frac{
 765731280722948607015073125 \pi^2}{34359738368}$ & 0.001643740 \\
 222120 & $\frac{86118466392025158318519831954180381}{326730875309916160} - \frac{
 917605934965893225554877375 \pi^2}{34359738368}$ & 0.000736923 \\
 222210 & $\frac{92310180650256200098982022759110013}{326730875309916160} - \frac{
 983579633627715621163792125 \pi^2}{34359738368}$ & 0.001813282 \\
 313110 & $\frac{9419406188801653071324696199909185}{65346175061983232} - \frac{
 501826343687610010797853125 \pi^2}{34359738368}$ & 0.000925144 \\
 322002 & $\frac{1461010403490121872831622811542911}{14205690230865920} - \frac{
 358047850704790643389117125 \pi^2}{34359738368}$ & 0.000588477 \\
 322020 & $\frac{1461010403490121872831622811542911}{14205690230865920} - \frac{
 358047850704790643389117125 \pi^2}{34359738368}$ & 0.000588477 \\
 322110 & $\frac{6000684679046793008540889210678705}{32673087530991616} - \frac{
 639382481602359423218915625 \pi^2}{34359738368}$ & 0.001780597 \\
 331002 & $\frac{254609784811217008522112965920873}{3843892650704896} - \frac{
 230597153568305033753368125 \pi^2}{34359738368}$ & 0.001457405 \\
 331020 & $\frac{254609784811217008522112965920873}{3843892650704896} - \frac{
 230597153568305033753368125 \pi^2}{34359738368}$ & 0.001457405 \\
 331110 & $\frac{454435716464909988842115675653445}{3843892650704896} - \frac{
 411577201458617736657740625 \pi^2}{34359738368}$ & 0.003076417 \\
 421002 & $\frac{198029832630946562183865640160679}{3843892650704896} - \frac{
 179353341664237248474841875 \pi^2}{34359738368}$ & 0.001133537 \\
 421020 & $\frac{198029832630946562183865640160679}{3843892650704896} - \frac{
 179353341664237248474841875 \pi^2}{34359738368}$ & 0.001133537 \\
 421110 & $\frac{353450001694929991321645525508235}{3843892650704896} - \frac{
 320115601134480461844909375 \pi^2}{34359738368}$ & 0.002392769 \\
\end{longtable}
\end{center}

Next, we come to the function ${\cal E}_{l_1m_1,\cdots,l_6m_6}$ given in Eq.~(\ref{calE}). Since this function involves the right-handed polarization vector $\hat{\epsilon}_{R}(\hat{k})$ of the photon field, we shall introduce the construction of this polarization vector and illustrate in some details how to work out the related quantities.

Let $\hat{k}$ be the unit vector which represents the propagation direction of the photon. Together with the polarization vectors $\hat{\varepsilon}_{1}(\hat{k})$ and $\hat{\varepsilon}_{2}(\hat{k})$, they form an orthonormal basis. To construct the polarization vectors, we need to choose an arbitrary constant unit vector $\hat{u}_{0}$. The polarization vectors can then be defined as
\begin{eqnarray}
    \hat{\varepsilon}_{1}(\hat{k})=\hat{\varepsilon}_{2}(\hat{k})\times\hat{k}\hskip 20pt ,\hskip 20pt \hat{\varepsilon}_{2}(\hat{k})=\frac{\hat{u}_{0}\times\hat{k}}{|\hat{u}_{0}\times\hat{k}|}.
\end{eqnarray}
While circular polarization vectors are defined as
\begin{eqnarray}
    \hat{\varepsilon}_{R,L}(\hat{k})=\frac{1}{\sqrt{2}}\,[\hat{\varepsilon}_{1}(\hat{k})\pm i\hat{\varepsilon}_{2}(\hat{k})],
\end{eqnarray}
with $\hat{\varepsilon}_{R,L}^{*}(\hat{k})=\hat{\varepsilon}_{L,R}(\hat{k})=\hat{\varepsilon}_{R,L}(-\hat{k}).$

In our calculations, we have to consider the inner product of the polarization vectors of two different photons, $\hat{\varepsilon}_{R}(\hat{k})\cdot\hat{\varepsilon}_{R}(-\hat{k}')$. Using the definitions above, we have
\begin{eqnarray}\label{2epsilon}
   && \hat{\varepsilon}_{R}(\hat{k})\cdot\hat{\varepsilon}_{R}(-\hat{k}')\nonumber\\
   &=&\frac{1}{2|\hat{u}_{0}\times\hat{k}||\hat{u}_{0}\times\hat{k}'|}\bigg\{\bigg[1+\hat{k}\cdot\hat{k}'-(\hat{u}_{0}\cdot\hat{k})(\hat{u}_{0}\cdot\hat{k}')-(\hat{u}_{0}\cdot\hat{k})^{2}-(\hat{u}_{0}\cdot\hat{k}')^{2}\nonumber\\
   &&\hskip 120pt +(\hat{u}_{0}\cdot\hat{k})(\hat{u}_{0}\cdot\hat{k}')(\hat{k}\cdot\hat{k}')\bigg]+i(\hat{u}_{0}\cdot(\hat{k}\times\hat{k}'))\bigg[\hat{u}_{0}\cdot\hat{k}+\hat{u}_{0}\cdot\hat{k}'\bigg]\bigg\}.\nonumber\\
\end{eqnarray}
To go further, we can get more explicit expressions by specifying the constant unit vector $\hat{u}_{0}$. Here, it is convenient for our purpose to do the integrals over solid angles of the $k$'s by taking $\hat{u}_{0}=(0,0,1)$. With $\hat{k}=(\sin\theta\cos\phi,\sin\theta\sin\phi,\cos\theta)$, the inner product of the two polarization vectors can be simplified to
\begin{eqnarray}\label{polprod}
    &&
    \hat{\varepsilon}_{R}(\hat{k})\cdot\hat{\varepsilon}_{R}(-\hat{k}')\nonumber\\
    &=&
    \frac{1}{2}\left[\cos(\phi-\phi')(1+\cos\theta\cos\theta')-i(\cos\theta+\cos\theta')\sin(\phi-\phi')+\sin\theta\sin\theta'\right].
\end{eqnarray}
With this simplification, the integrations over the angles of $\hat{k}$, $\hat{k}'$, and $\hat{k}''$ can be evaluated for particular sets of $l$ and $m$. 
To facilitate this, we make use of the isotropic basis functions introduced in Eq.~(\ref{eqn:isofunc_4pcf}) to expand the product of polarization vectors of the integrand in Eq~(\ref{calE}) as
\begin{eqnarray}\label{epsilonexp}
    (\hat{\epsilon}_{R}(\hat{k})\cdot\hat{\epsilon}_{R}(-\hat{k}'))
    (\hat{\epsilon}_{R}(\hat{k}')\cdot\hat{\epsilon}_{R}(-\hat{k}''))(\hat{\epsilon}_{R}(\hat{k}'')\cdot\hat{\epsilon}_{R}(-\hat{k}))
    =\sum_{\ell,\ell',\ell''}\epsilon_{\ell\,\ell'\,\ell''}{\cal P}_{\ell\,\ell'\,\ell''}(\hat{k},\hat{k}',\hat{k}'')\,. \nonumber\\
\end{eqnarray}
From the orthogonality relation of the isotropic basis functions in Eq.~(\ref{isoortho}), the coefficient function $\epsilon_{\ell,\ell',\ell''}$ can be expressed as
\begin{eqnarray}
    \epsilon_{\ell\,\ell'\,\ell''}&=&\int\,d\hat{k}\,d\hat{k}'\,d\hat{k}''\,{\cal P}^{*}_{\ell\,\ell'\,\ell''}(\hat{k},\hat{k}',\hat{k}'')\nonumber\\
    &&\ \ \ \ (\hat{\epsilon}_{R}(\hat{k})\cdot\hat{\epsilon}_{R}(-\hat{k}'))
    (\hat{\epsilon}_{R}(\hat{k}')\cdot\hat{\epsilon}_{R}(-\hat{k}''))(\hat{\epsilon}_{R}(\hat{k}'')\cdot\hat{\epsilon}_{R}(-\hat{k}))\,.
\end{eqnarray}
From the symmetry properties of the polarization vector products above as well as those of the isotropic basis functions, we can see that the coefficients $\epsilon_{\ell,\ell',\ell''}$ are totally symmetric with respect to $\ell$, $\ell'$, and $\ell''$. By direct calculation, using the expression for the product of polarization vectors in Eq.~(\ref{polprod}), we have found that the only non-zero coefficients are listed as follows:
\begin{eqnarray}\label{epsiloncoef}
   && \epsilon_{000}=\frac{8\pi^{3/2}}{9}\ \ ,\ \ \epsilon_{110}=-\frac{4\pi^{3/2}}{3\sqrt{3}}\ \ ,\ \ \epsilon_{111}=\frac{\sqrt{2}\pi^{3/2}}{3}\ \ ,\nonumber\\
   &&\epsilon_{211}=\frac{\sqrt{2}\pi^{3/2}}{3\sqrt{3}}\ \ ,\ \ \epsilon_{220}=\frac{4\pi^{3/2}}{9\sqrt{5}}\ \ ,\ \ \epsilon_{221}=-\frac{\sqrt{2}\pi^{3/2}}{3\sqrt{5}}\ \ ,\ \ \epsilon_{222}=\frac{\sqrt{14}\pi^{3/2}}{45}\ \ .
   \end{eqnarray}

With this expansion of the products of polarization vectors, the function ${\cal E}_{l_1m_1,\cdots,l_6m_6}$ can now be worked out quite readily.
\begin{eqnarray}
    {\cal E}_{\ell _{1}m_{1}\cdots \ell _{6}m_{6}}&=&\sum_{\ell\ell'\ell''}\epsilon_{\ell\ell'\ell''}\int\,d\hat{k} d\hat{k'}d\hat{k''}\,{\cal P}_{\ell\,\ell'\,\ell''}(\hat{k},\hat{k}',\hat{k}'')\nonumber\\
    &&\hskip 50pt Y_{\ell _{1}m_{1}}^{*}(\hat{k})Y_{\ell _{2}m_{2}}(\hat{k})Y_{\ell _{3}m_{3}}^{*}(\hat{k'})Y_{\ell _{4}m_{4}}(\hat{k'})Y_{\ell _{5}m_{5}}^{*}(\hat{k''})Y_{\ell _{6}m_{6}}(\hat{k''}) \nonumber\\
    &=&\sum_{\ell\ell'\ell''}\epsilon_{\ell\ell'\ell''}\ (-1)^{\ell+\ell'+\ell''}\sum_{m\,m'\,m''}\six{\ell}{\ell'}{\ell''}{m}{m'}{m''}\int\,d\hat{k}\, Y_{\ell _{1}m_{1}}^{*}(\hat{k})Y_{\ell _{2}m_{2}}(\hat{k})Y_{\ell m}(\hat{k})\nonumber\\
    &&\int\,d\hat{k'}\,Y_{\ell _{3}m_{3}}^{*}(\hat{k'})Y_{\ell _{4}m_{4}}(\hat{k'})Y_{\ell' m'}(\hat{k}')\int\,d\hat{k''}\,Y_{\ell _{5}m_{5}}^{*}(\hat{k''})Y_{\ell _{6}m_{6}}(\hat{k''})Y_{\ell'' m''}(\hat{k}'')\,,\nonumber\\
\end{eqnarray}
where we have made use of the definition of the isotropic basis function in Eq.~(\ref{eqn:isofunc_4pcf}). The integrals with three spherical harmonics can be expressed using the Wigner-3j symbols again as indicated in Eq.~(\ref{YYYint}). Finally, we have
\begin{eqnarray}\label{coeffinal}
     {\cal E}_{\ell _{1}m_{1}\cdots \ell _{6}m_{6}}&=&\sum_{\ell\ell'\ell''}\epsilon_{\ell\ell'\ell''}\ \frac{(-1)^{\ell+\ell'+\ell''+m_{1}+m_{3}+m_{5}}}{8\pi^{3/2}}\sqrt{(2\ell+1)(2\ell'+1)(2\ell''+1)\prod_{i=1}^{6}(2\ell_{i}+1)}\nonumber\\
     &&\hskip 50pt \six{\ell}{\ell_1}{\ell_2}{0}{0}{0}\six{\ell'}{\ell_3}{\ell_4}{0}{0}{0}\six{\ell''}{\ell_5}{\ell_6}{0}{0}{0}\nonumber\\
     &&\sum_{m\,m'\,m''}\six{\ell}{\ell'}{\ell''}{m}{m'}{m''}\six{\ell}{\ell_1}{\ell_2}{m}{-m_1}{m_2}\six{\ell'}{\ell_3}{\ell_4}{m'}{-m_3}{m_4}\six{\ell''}{\ell_5}{\ell_6}{m''}{-m_5}{m_6}.\nonumber\\
\end{eqnarray}
In this form, together with the values of the coefficients given in Eq.~(\ref{epsiloncoef}), the function ${\cal E}_{l_1m_1,\cdots,l_6m_6}$ can be evaluated just by summing over various Wigner-3j symbols. 

\section{A toy inflation model}
\label{sec:example}
\begin{figure}[htp]
\centering
\includegraphics[width=0.8\textwidth]{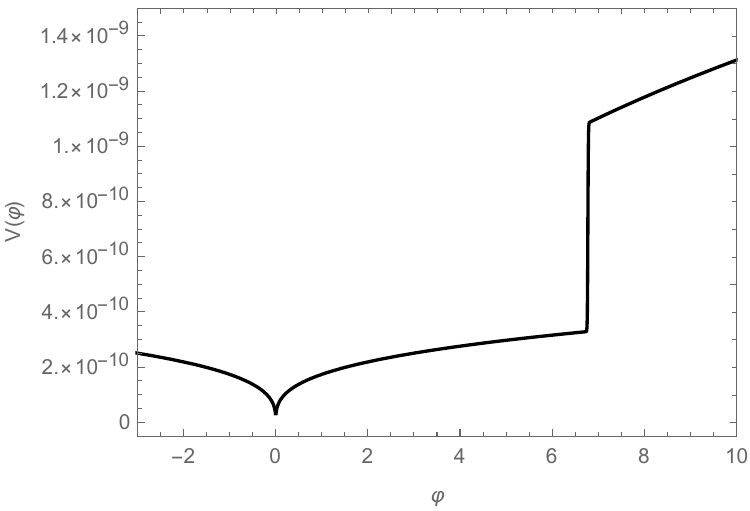}
\caption{Inflaton potential $V(\varphi)$ with $\mu_1=1.74 \times 10^{-10}$,
$\mu_2=8.7 \times 10^{-10}$, $C=1000$, $A=7.55 \times 10^{-10}$, $s=100$, and
$d=6.775$. Here and in the following figures, we have set the reduced Planck mass $M_{P}=1$.}
\label{Vphi}
\end{figure}

\begin{figure}[htp]
\centering
\includegraphics[width=0.8\textwidth]{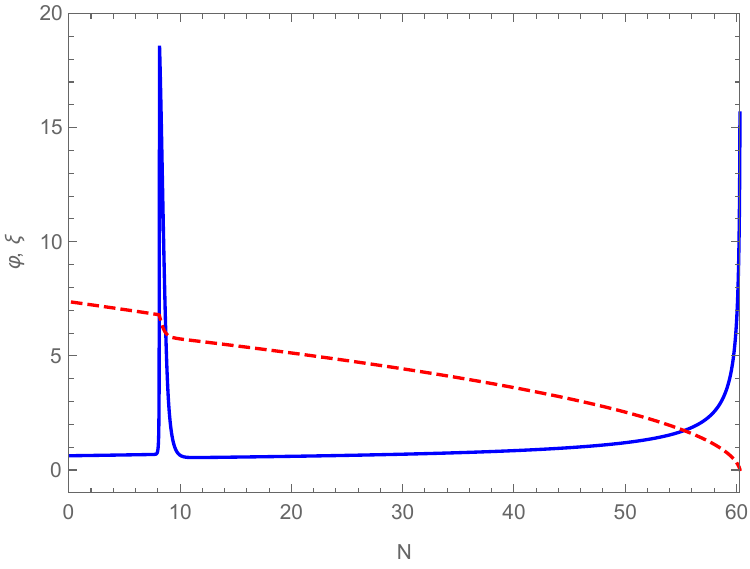}
\caption{Evolution of $\varphi$ (dashed line) and $\xi$ (solid line),
with $\varphi_0=7.3754$, $(d\varphi/dt)_0=-1.3095 \times 10^{-6}$, $H_0=1.9429 \times 10^{-5}$, and $\alpha=18.5$. Inflation lasts for $60.29$ e-foldings.}
\label{phixi}
\end{figure}

\begin{figure}[htp]
\centering
\includegraphics[width=0.8\textwidth]{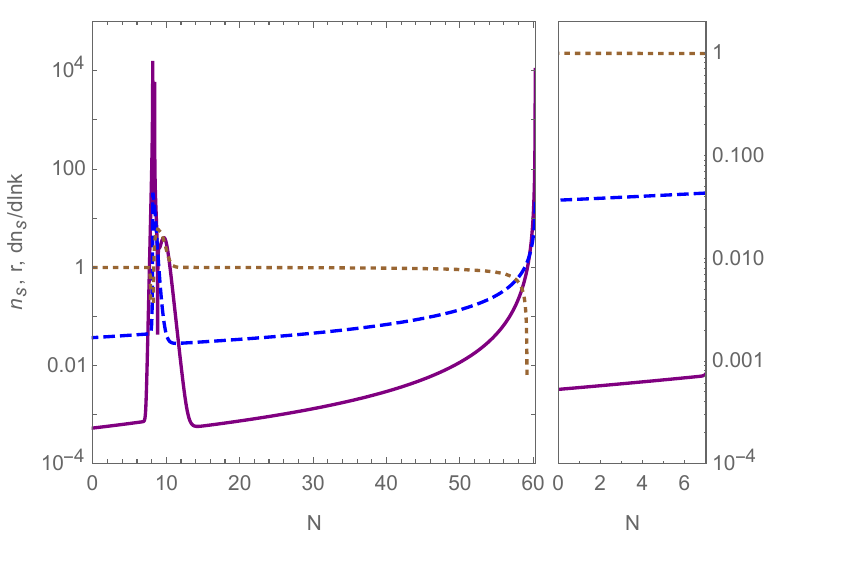}
\caption{Evolution of $n_s$ (dotted line), $r$ (dashed line), and $dn_s/d\ln k$ (solid line). The right panel zooms in on the first $7$ e-foldings, drawn with $|dn_s/d\ln k|$.}
\label{index}
\end{figure}

\begin{figure}[htp]
\centering
\includegraphics[width=0.8\textwidth]{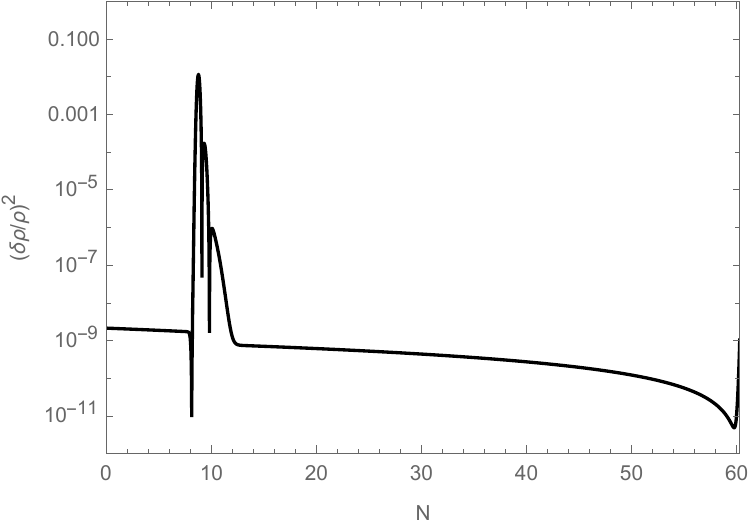}
\caption{Density power spectrum.}
\label{spectrum}
\end{figure}

Following the numerical method in Refs.~\cite{CLN16,CLN18}, which takes into account the backreaction of the photon production to inflation, we have worked out a toy model with the inflaton potential given by
\begin{equation}
V\left( \varphi \right) = \frac{\mu_1}{C}\left[ \left( 1+C^6 \varphi^2 \right)^\frac{1}{6} - 1 \right] + \left[ \left( \mu_2 - \mu_1 \right) \left( \varphi^\frac{1}{3} - d^\frac{1}{3} \right) + A \right] \frac{1}{2} \left( 1 + \tanh \left[ \left( \varphi - d \right) s \right] \right)\,.
\end{equation}
The parameters used in the model and the results are shown in Figs.~\ref{Vphi}-\ref{spectrum}.
The amplitude of the density power spectrum $(\delta\rho/\rho)^2$, 
the scalar spectral index $n_s$, the tensor-to-scalar ratio $r$, and the scalar spectral index running $dn_s/d\ln k$,
are predicted to be consistent with the CMB measurements: 
$\Delta_\zeta^2\simeq 2.1\times 10^{-9}$, $n_s\simeq 0.97$, $r<0.036$, 
and $|dn_s/d\ln k|<0.013$~\cite{planckparameter20,BICEP/Keck21}, respectively, 
at the CMB or cosmological scales that correspond to about the first $7$ e-foldings of inflation.

\end{document}